\documentclass{aa}  

\usepackage{graphicx}
\usepackage{txfonts,textcomp}
\usepackage{hyperref}
\usepackage[dvipsnames]{xcolor}
\usepackage{booktabs,multirow,tabularx}
\newcommand{\makecell}[2][@{}c@{}]{\begin{tabular}{#1}#2\end{tabular}}

\makeatletter
\renewcommand*\aa@pageof{, page \thepage{} of \pageref*{LastPage}}
\makeatother

\usepackage[desactivate]{linenoaa}

\begin{document}

   \title{Impact of far-side structures observed by Solar Orbiter on coronal and heliospheric wind simulations}

   \author{B. Perri\inst{1}
          \and
          A. Finley\inst{1}
          \and 
          V. Réville\inst{2}
          \and
          S. Parenti\inst{3}
          \and
          A. S. Brun\inst{1}
          \and
          A. Strugarek\inst{1}
          \and
          É. Buchlin\inst{3}
          }

   \institute{Dept. d'Astrophysique/AIM, CEA/IRFU, CNRS/INSU, Universit\'e Paris et Paris-Saclay, \\ 91191 Gif-sur-Yvette Cedex, France
         \and
             IRAP, Université Toulouse III - Paul Sabatier, CNRS, CNES, Toulouse, France
         \and
             Université Paris-Saclay, CNRS, Institut d’Astrophysique Spatiale, 91405, Orsay, France
             }

   \date{Received ??; accepted ??}

 
  \abstract
   {Solar Orbiter is a new space observatory that provides unique capabilities to understand the heliosphere. In particular, it has made several observations of the far-side of the Sun and therefore provides unique information that can greatly improve space weather monitoring.}
   {In this study, we aim to quantify how the far-side data will affect simulations of the corona and the interplanetary medium, especially in the context of space weather forecasting.}
   {To do so, we focused on a time period with a single sunspot emerging on the far-side in February 2021. We used two different input magnetic maps for our models: one that includes the far-side active region and one that does not. We used three different coronal models typical of space weather modeling: a semi-empirical model (potential field source surface or PFSS) and two different magnetohydrodynamic models (Wind Predict and Wind Predict-AW). We compared all the models with both remote sensing and in situ observations in order to quantify the impact of the far-side active region on each solution.}
   {We find that the inclusion of the far-side active region in the various models has a small local impact due to the limited amount of flux of the sunspot (at most 8\% of the total map flux), which leads, for example, to coronal hole changes of around 7\% for all models. Interestingly, there is a more global impact on the magnetic structure seen in the current sheet, with clear changes, {for example,} in the coronal hole boundaries {visible} in extreme ultra-violet (EUV) on the western limb, which is opposite to the active region and the limb most likely to be connected to Earth. For the Wind Predict-AW model, we demonstrate that the inclusion of the far-side data improves both the {structure} of the streamers and the connectivity to the spacecraft.}
   {In conclusion, the inclusion of a single far-side active region may have a small local effect with respect to the total magnetic flux, but it has global effects on the magnetic structure, and thus it must be taken into account to accurately describe the Sun-Earth connection. The flattening of the heliospheric current sheet for all models reveals that it causes an increase of the source surface height, which in return affects the open and closed magnetic field line distributions.}


   \maketitle
%

\section{Introduction}

Space weather forecasting is usually performed using a chain of models that connect the solar surface up to the near-Earth space environment, such as in the Virtual Space Weather Modeling Center (VSWMC) for Europe \citep{Poedts2020_vswmc}, the Solar Wind Modeling Framework (SWMF) for the United States \citep{Odstrcil2003, Toth2012}, {or the SUSANOO model in Japan \citep{Shiota2014}}. The first element of this chain is the coronal model, which typically takes as input some measurements {between 1 and 3 solar radii} (magnetic field in most cases but sometimes density or white-light images are used, as in \cite{Reville2023}), and it provides as output the state of the solar corona near 20 solar radii. This output is then used to drive heliospheric models that propagate the solar wind along with transients, such as coronal mass ejections (CMEs), out to the Earth's orbit at one astronomical unit. The coronal part is one of the most crucial elements of this chain of modeling, not only because it is the source of the solar wind structures that will influence the rest of the chain but also because it is one of the regions that contains the most challenging physics. Different modeling can thus drastically change the propagation of transients and the description of the space weather situation at Earth \citep{Samara2021}.

To achieve an accurate description of the solar wind and magnetic field in the solar corona, there are many different possible approaches. The least resource-intensive technique is to use empirical extrapolations such as the WSA method \citep[Wang-Sheeley-Arge, see ][]{Wang1990,Arge2000}, which computes the wind speed based on an inverse relationship with the flux tube expansion from a static force-free extrapolation of the coronal magnetic field.  
Capturing the non-force free features in the corona requires the use of magneto-frictional models \citep{Pomoell2019} or even full-on magnetohydrodynamic (MHD) models. Even for MHD modelling, there is a wide range of approximations being used for the heating and acceleration of the wind. One-dimensional wind models have proved to be interesting for their capacity of including realistic physics at reasonable computational costs \citep{Lionello2001, Suzuki2005, Grappin2010,Pinto2017}. Three-dimensional MHD models have been developed and improved over the years with systematic comparisons to both in situ and remote-sensing data for validation, such as AWSoM \citep{vanderHolst2010, vanderHolst2014, Sachdeva2019}, Wind Predict \citep{Reville2015a, Perri2018, Reville2020, Parenti2022}, or COCONUT \citep{Perri2022, Perri2023, Kuzma2023}. Other models can be even more sophisticated, but this comes at a computational cost such that they can no longer be used in operational space weather forecasting \citep{Mikic2018}. Some models even go beyond the single fluid approximation by taking into account the multi-species nature of the solar wind \citep{Usmanov2014, Usmanov2016, Usmanov2018, Chhiber2021}. Yet other models have focused on boundary conditions to provide the most physical conditions for time-dependent models \citep{Wu2006}, such as in \cite{Yalim2017} and \cite{Singh2018}. 
(For a somewhat exhaustive list of solar wind models used for space weather forecasting, please see \cite{Reiss2022unifying}.) 

In all the aforementioned approaches, coronal and heliospheric simulations are data driven and therefore suffer when given incomplete data as an input, the most systematic of which is a lack of far-side observations of the Sun. As a result, synoptic maps based on solar data used to initialize numerical simulations are never synchronic, that is, they need to include data from different time periods in order to cover the full 360 degrees of the Sun in longitudes. Using asynchronous maps is often a reasonable approximation when solar activity is close to minimum, as there is little evolution during a solar rotation. However, at maximum solar activity, this begins to cause serious issues in the input data, as active regions (ARs) can emerge on the far-side without being taken into account into the synoptic maps. New developments in helioseismology are currently trying to address this issue, but synoptic products based on this method are still in development \citep{heinemann2023farm}. The synchronicity of solar observations was solved briefly by the two STEREO spacecraft that were spread around the Sun in order to have full coverage of solar activity. However, due to the loss of STEREO-B in 2014 and the migration of STEREO-A back towards the Earth, there were no far-side observations until recently with the launch of Solar Orbiter in 2020 \citep{Muller2020}. Solar Orbiter is currently the only observatory that observes the Sun {with multiple multi-wavelength remote sensing instruments} far outside the Sun-Earth line, which means that there are moments when Solar Orbiter is observing the far-side of the Sun, thus fully complementing the observations from the point of view of the Earth.

The impact of including far-side ARs in operational space weather modelling is thus so far untested. So in order to anticipate the value of upcoming data products from Solar Orbiter, which incorporate far-side observations, we performed numerical experiments. As a first step, we selected a case {when solar activity is at a minimum}, which allowed us to study a singular AR and to already quantify its impact in this simple configuration. We therefore examined a period of time during Solar Orbiter's cruise phase in which an AR emerged on the far-side of the Sun but in view of Solar Orbiter. The aim is to quantify how the inclusion of this region in the input synoptic map influences the resulting coronal and heliospheric wind simulation, which in turn are likely to affect CME propagation and space weather forecasts at Earth.

The article is organized as follows. In Section~\ref{sec:methodology}, we describe our methodology for taking into account the far-side AR in the driving of the coronal models. In Section~\ref{sec:codes}, we briefly describe the models used in this study, that is, the potential field source surface (PFSS) extrapolation and the two variants of our 3D MHD solar corona model used in this study, namely, Wind Predict (WP) and Wind Predict with Alfvén Waves (WP-AW).
{Detailed descriptions of each coronal model can be found in Appendix~\ref{app:codes}. We present a comparison of our results with the remote-sensing observations to constrain the impact of the AR on the inner corona, focusing on closed and open magnetic structures in Section~\ref{sec:results_wl} and Section~\ref{sec:results_ch}, respectively. In Section~\ref{sec:results_insitu}, we present a comparison of our results with the in situ observations to constrain the impact of the AR on the inner heliosphere. To conclude, our findings are discussed in Section~\ref{sec:discussion}, where we offer explanations for the differences observed between the models} before discussing future perspectives to this work in Section \ref{sec:conclusion}.

\section{Methodology}
\label{sec:methodology}

\subsection{Selection of the event with Solar Orbiter data}

\begin{figure*}[!ht]
    \centering
    \includegraphics[width=\textwidth]{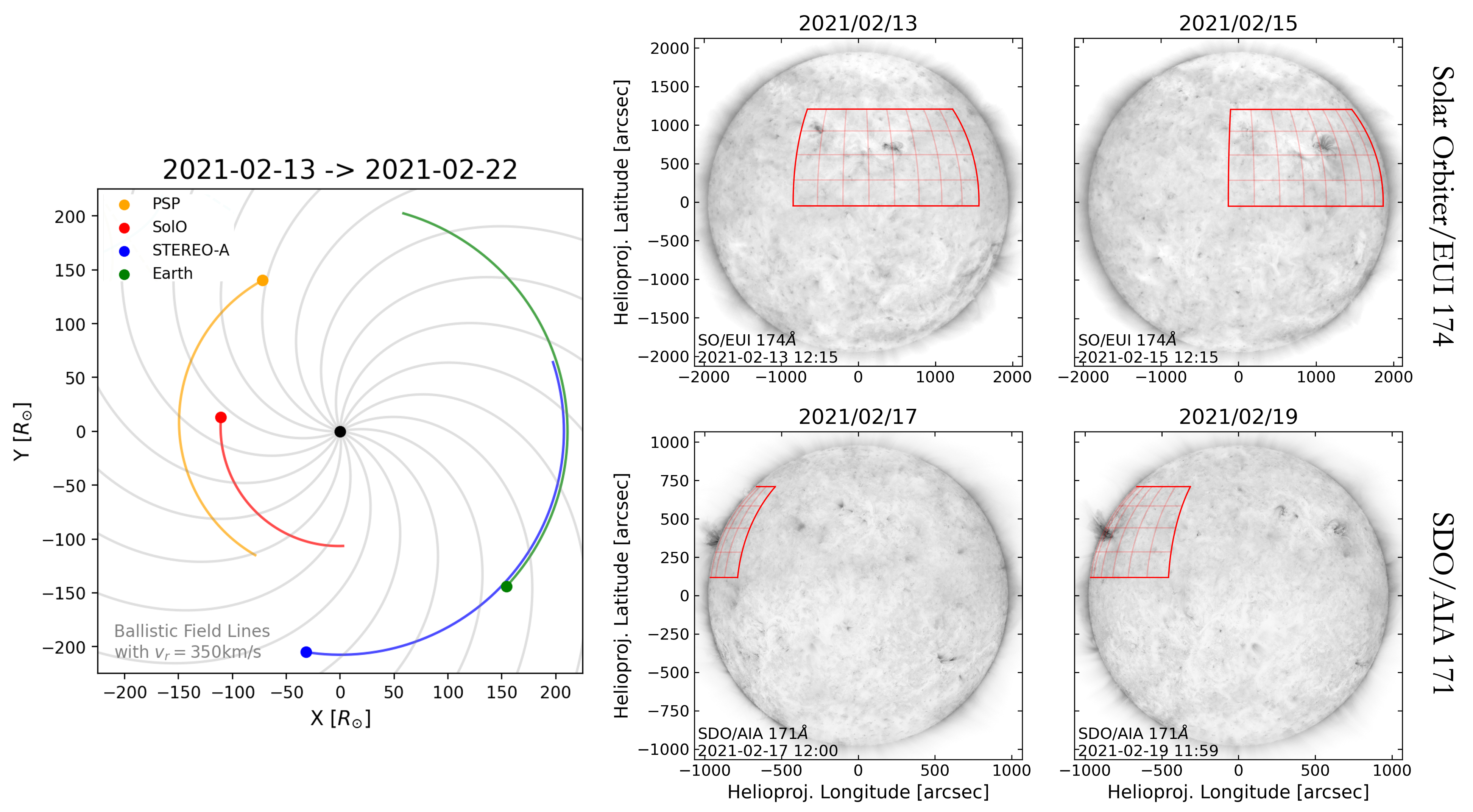}
    \caption{Observations of AR12803 during February 2021. The left panel shows the orbital configuration of the various satellites from February 13 to February 22, 2021 in the Carrington frame synchronized with the surface of the Sun. Solar Orbiter was covering the far-side of the Sun during this period. The upper-right panel shows observations by Solar Orbiter/EUI 174\,\AA\ between February 13 and 15, 2021, where one can see the emergence of the AR. The lower-right panel shows observations by SDO/AIA 171\,\AA\ between February 19 and 21, 2021 when the AR finally entered the Earth field of view. In both cases, the AR of interest is located inside the red grid.}
    \label{fig:hmi_phi_obs}
\end{figure*}

The goal of this study is to quantify the impact of an emerging far-side AR observed by Solar Orbiter on data-driven 3D simulations of the heliosphere. 
Such an event happened in early 2021 during Solar Orbiter cruise phase. On February 13, 2021, an AR emerged on the far-side of the Sun, and was observed by Solar Orbiter EUI instrument (Extreme Ultraviolet Imager, see \cite{Rochus2020}) while the satellite was in opposition with the Earth. The AR reached its maximum amplitude on February 15 and finally rotated in the field of view of the Earth on February 20, where it was labeled AR12803. This sequence is illustrated in Figure~\ref{fig:hmi_phi_obs}. The left panel shows the orbital configuration of the various satellites from February 13 to February 22, 2021, as viewed from above in the Sun's rotating frame of reference. Solar Orbiter (in red) was covering the far-side of the Sun (Earth in green) during this period. The upper-right panel shows observations by Solar Orbiter/EUI 174\,\AA\ between February 13 and 15, 2021, where the emergence of the AR inside the red grid can be seen. The lower-right panel shows observations by the Atmospheric Imaging Assembly (AIA) {\citep{Lemen2012}} onboard the Solar Dynamics Observatory (SDO) {\citep{Pesnell2012}} with the 171\,\AA\ channel, between February 19 and 21, 2021, when the AR finally entered the Earth field of view.

We focused specifically on this period because it was close to solar cycle 25 minimum of activity (December 2020). {This means that there were no major erupting structure that could change the global configuration of the heliosphere over the studied period between February 15 and 22, 2021. Thus, the underlying assumption of our quasi-static models, that the photospheric magnetic field does not evolve much and can be reproduced using only a steady-state solution, is reasonable.} 

\subsection{Input magnetic maps selection}

In order to quantify the impact of this far-side emerging AR, our approach in this paper is to compare coronal models initiated with different input data. 
We first use a synoptic map describing the magnetic field at the surface of the Sun before the emergence (so without the AR), and compare it with simulations driven by a second synoptic map that does include the far-side region after its emergence.

\begin{figure*}[htbp]
    \centering
    \includegraphics[width=\textwidth]{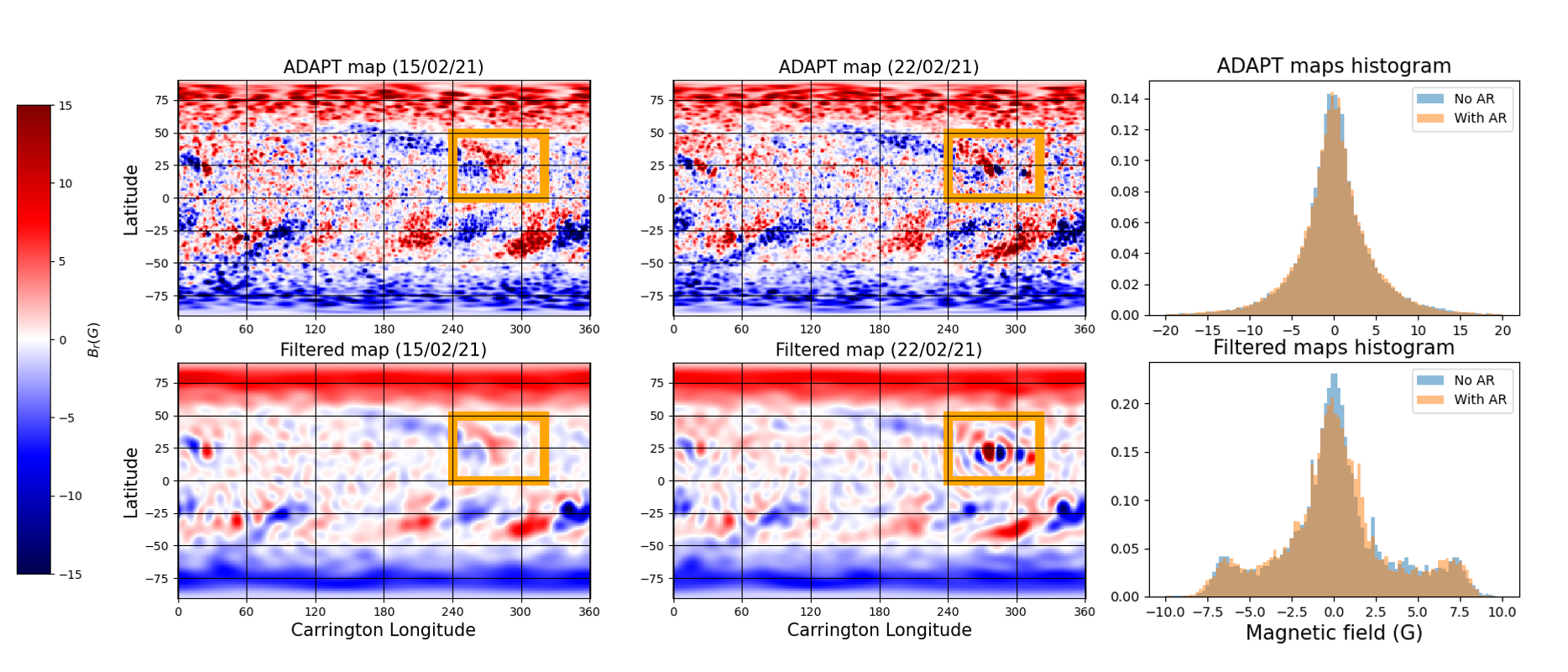}
    \caption{Analysis of the magnetic input maps selected for the study. The left panels show the maps for February 15, 2021, and the middle panels the maps for February 22, 2021. The top row shows the original maps (both GONG-ADAPT maps in Carrington longitudes), while the bottom row shows the filtered maps used to initialize the simulations (spherical harmonics reconstruction up to $\ell_{max}=30$). The far-side AR location is marked with an orange box. In the right panel is a histogram analysis of the radial magnetic field distribution for original (top) and filtered maps (bottom), without (blue) and with (orange) the far-side AR.}
    \label{fig:adapt_maps}
\end{figure*}

Solar Orbiter has a magnetograph dedicated to measuring the surface solar magnetic field called PHI (Polarimetric and Helioseismic Imager; see \cite{Solanki2020}). The ideal way to assess the impact of the far-side AR would be to use a magnetic map relying on old asynchronous data and a second map with updated PHI measurements.
Unfortunately, since the event selected was during Solar Orbiter's cruise phase and since there are no official HMI/PHI or GONG/PHI merged products yet, we cannot directly use the Solar Orbiter data. Instead, we selected two GONG-ADAPT magnetic maps\footnote{\url{https://gong.nso.edu/adapt/maps/}} at specific dates to emulate the emergence of the far-side AR. {This is shown in Figure~\ref{fig:adapt_maps}.} The first map is the ADAPT map from February 15, 2021, at 12:00 UTC, which was made using data assimilation from previous observations, but only from the Earth field of view. This means that, although the AR was emerged at that time, it will not be visible on the magnetic map. The other input data we chose was again an ADAPT map, but from February 22 at 22:00 UTC. At this date, the AR was visible from Earth, and thus is visible also in the map (around 280 degrees of longitude and 25 degrees of latitude, region marked with an orange box on {Figure~\ref{fig:adapt_maps}}). It was the main change in the magnetic field configuration over this period of time, so that the rest of the map outside the AR is very similar between both maps. For each ADAPT map, we have used only the first realization (number 0 in the FITS file). Both maps are in Carrington frames, which means that the longitudes only depend on the current Carrington rotation. For clarity, we also show on the bottom panel the filtered maps that we use to initialize the various models (spherical harmonics reconstruction of the field with $\ell_{max}=30$). This reconstruction filters out the smallest and most intense structures, but we can still see clearly the emergence of the far-side AR of interest. As can be seen in Figure~\ref{fig:adapt_maps}, we could thus use the map from February 22 as if it were an updated map of February 15 with the far-side AR included. The AR would be located at the same Carrington longitude corresponding to the far-side, and we could then run our simulations and see how the inner heliosphere is affected.

\subsection{Input magnetic maps analysis}

Before running the simulations, we can already quantify the impact of the far-side AR on the input maps. This is the purpose of the right panels of Figure~\ref{fig:adapt_maps}, where we plot the histogram distribution (with 100 bins) of the radial magnetic field in the previous maps. The top panel is for the original maps, while the bottom panel is for the filtered maps. In blue, we show the distribution without the far-side AR, and in orange the distribution with the AR included. 
We observed that the general distribution is as expected: a Gaussian centered around zero. For the original maps, the wings extend up to 20 G, while for the filtered maps they extend only up to 10 G (as a result of the filtering of the most intense structures). We observed that the distribution is also slightly altered, with two bumps around $\pm7.5$ G that may bring it closer to a Bessel function than a Gaussian. Our focus is however on the effect of the inclusion of the AR. In both histograms, one can see that it depopulates the levels of the smallest values around zero and increases the values in the wings in an almost uniform way. Therefore, we observed a global increase of the intensity of the magnetic field but no specific peak associated with one value.
We have also computed the unsigned flux of the radial magnetic field in the region of interest, shown by an orange rectangle on Figure~\ref{fig:adapt_maps}. We find that the inclusion of the AR in the original map leads to an increase in flux in the region of interest of a factor of 1.95 ($2.79 \, 10^{22}$ Mx vs. $1.4 \, 10^{22}$ Mx). For the filtered maps, the factor is very similar with 2.11 ($1.34 \, 10^{22}$ Mx vs. $6.4 \, 10^{21}$ Mx). Though the filtering of the map impacts the general flux of the map and the AR with a factor two of reduction for both, its effect is thus negligible for the relative increase of the flux. Compared to the total unsigned flux of the original map, the flux in the AR represents only 8.3\% (total unsigned flux of $3.4 \, 10^{23}$ Mx). For the filtered map, this value goes down to 4\% of the total unsigned flux ($2.4 \, 10^{23}$ Mx). This gives us an estimate of the order of magnitude of the changes we can expect if we have a linear response from the corona.

\section{Numerical models}
\label{sec:codes}

\subsection{Parameters of models}

In this study, we use three different models, the analytical method called PFSS extrapolation and two different versions of the WP {MHD} model: the polytropic version of WP and the Alfvén wave-driven version, WP-AW. These 3D MHD coronal models are based on the multi-physics PLUTO code \citep{Mignone2007}.

The PFSS is the fastest method used for coronal forecasting {(converges in tens of minutes)}, but is {only} semi-empirical \citep{schrijver2003photospheric}. 

The WP model was first introduced and described for star-planet interactions in \cite{Strugarek2014}, and for stellar applications in \cite{Reville2015a, Reville2015b}. It has since been adapted specifically for the solar case \citep{Reville2017} and for spherical geometry \citep{Perri2018}. It has also been used as a reference case to calibrate other wind models \citep{Perri2022}. WP is faster to run {than WP-AW (it converges in less than 6 hours on 200 cores)} and is thus an operational space weather forecasting model,\footnote{\url{https://swe.ssa.esa.int/kul-cmpa-federated}} but is less physically realistic {than WP-AW}.

The WP-AW model has been validated against both in situ \citep{Reville2020, Reville2021} and remote-sensing observations \citep{Parenti2022}. {It has already been improved with new features, such as the inclusion of a realistic transition region \citep{Reville2021}, \citep{Reville2022}.}

{We would like to stress that none of these models have been fine-tuned for this specific case study in this work. Our approach is to use them in a set-up as close as operational as possible in order to both understand how the magnetic connectivity is modified by adding a new far side AR as well as estimate the impact of these potential modifications for space weather forecasting. The only adjustment made was for the WP model, where the polytropic index $\gamma$ was set to 1.03 instead of 1.05 in order to guaranty a better opening of the equatorial coronal holes.}

{Because these models are well established, we focus here only on their main parameters relevant to this study, which can be found in Table~\ref{tab:param_models}. For each model, we specify the input parameters related to the input magnetic map (cut-off harmonics $\ell_{max}$ and multiplying factor $\mu$). We also specify parameters regarding the simulation domain and the grid resolution. Finally, we specify the parameters that are characteristic of each model. We end with the main numerical methods used, when it is relevant.}

\begin{table*}[!ht]
    \centering
    \begin{tabular}{c||c|c|c}
        Input parameters & PFSS & WP & WP-AW \\ \hline
        $\ell_{max}$ (cut-off harmonics of input map) & 30 & 30 & 25 \\
        $\mu$ (multiplying factor of input map) & 1 & 0.3 & 1 \\
        Domain ($r[R_\odot]\times\theta[\rm{rad}]\times\varphi[\rm{rad}]$) & $[0,2.5]\times[0,\pi]\times[0,2\pi]$ & $[0,30]\times[0,\pi]\times[0,2\pi]$ & $[0,30]\times[0.05,3.09]\times[0,2\pi]$ \\
        Grid ($r\times\theta\times\varphi$) & $12\times92\times184$ & $172\times86\times172$ & $182\times160\times320$\\
        Specific parameters & $R_{ss}=2.5\,R_\odot$ & $\gamma=1.03, \frac{c_s}{v_{esc}}=0.26$ & $F_h=2\, 10^4 \, \rm{cgs}, \kappa_0=9\,10^{-7} \, \rm{cgs},$ \\
         & & & $\delta v=12\rm{km/s}, T_c=2\,10^4\,K$ \\
        Numerical methods & N/A & HLLD, divergence cleaning & HLL, constrained transport \\
    \end{tabular}
    \caption{{Table summarizing the main modeling parameters for the three different models tested in this work (PFSS, WP, and WP-AW). For each model, we specify the input parameters related to the input magnetic map (cut-off harmonics $\ell_{max}$ and multiplying factor $\mu$). We also specify parameters regarding the simulation domain and the grid resolution. Finally, we specify the parameters that are characteristic of each model. We end with the main numerical methods used, when it is relevant. For a more precise definition of each parameter, please see Appendix~\ref{app:codes}.}}
    \label{tab:param_models}
\end{table*}

{From this table, we can make a few remarks. The filtered magnetograms where reconstructed with an $\ell_{max}$ of 25 for WP-AM because it is the best compromise between accuracy and speed of computation, with no major structure being erased. The WP model needs a multiplying factor of 0.3 applied to the input magnetic map to stay within its robustness parameter space, but once again this has very little effect on the realism of the global magnetic structure. The PFSS model only computes the configuration up until the $R_{ss}$ because after that the magnetic field lines are simply forced to be radial. A detailed description of the models' underlying physics and numerics can be found in Appendix~\ref{app:codes} for further details and explanations.}

\subsection{Overview of the different models}
\label{subsec:results_overview}

\begin{figure*}[htbp]
    \centering
    \includegraphics[width=\textwidth]{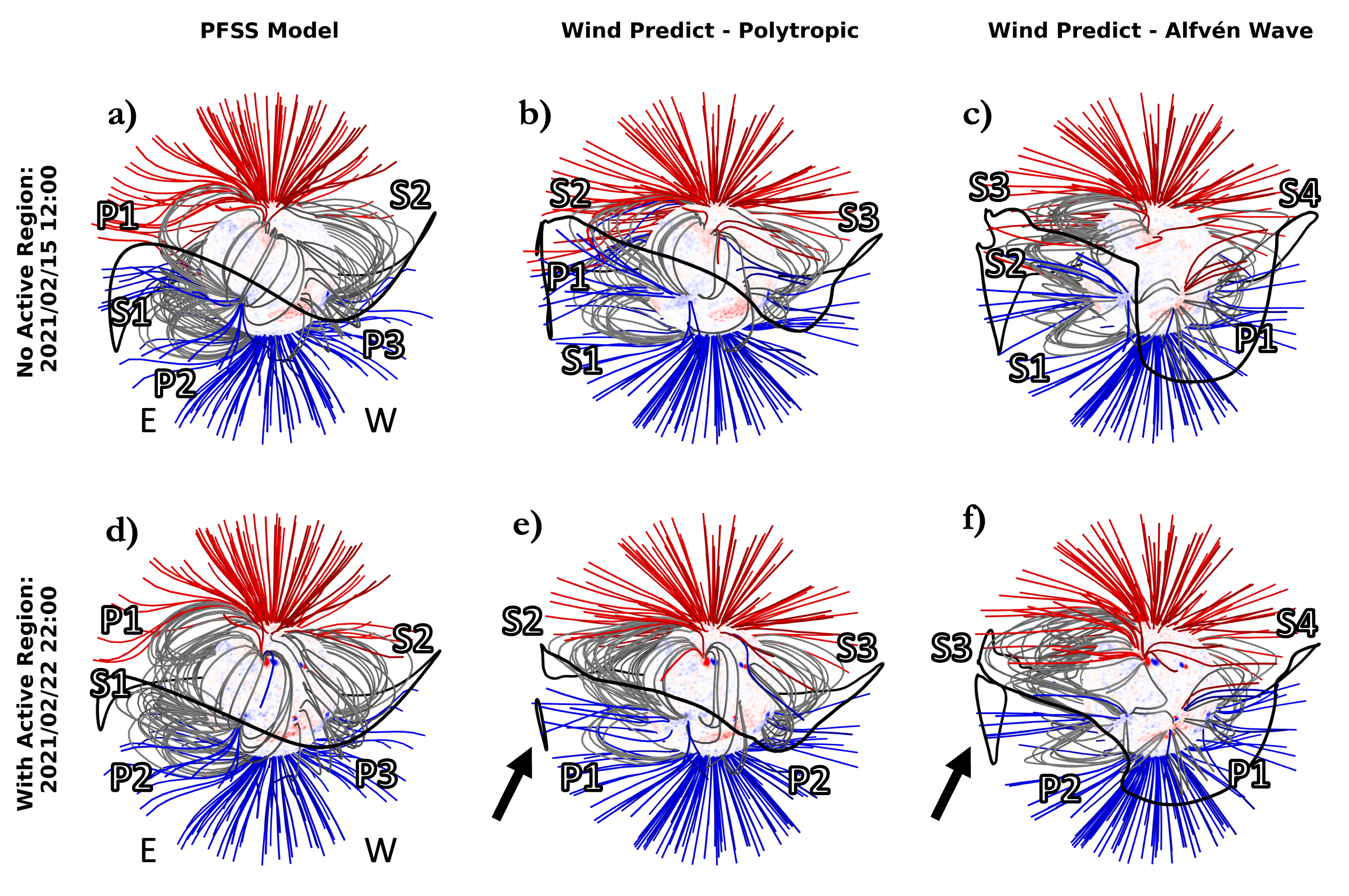}
    \caption{Three-dimensional overview of the magnetic field structure derived from the various models without (top row) and with the far-side AR (bottom row). The left panels show the PFSS results, the middle panels the WP results and the right panels the WP-AW results. Magnetic field lines are gray when closed, red when open with a positive polarity and blue when open with a negative polarity. The HCS is also shown as a continuous black line around the Sun. We indicate the location of the eastern and western limbs with the letter E and W respectively. {We also indicate whether a streamer is a helmet streamer contributing to the HCS (S) or a pseudo-streamer (P) \citep{Wang2007}, and we show their number. Arrows point out HCS structures that break apart from the main line.}}
    \label{fig:magnetic_overview}
\end{figure*}

We can have a first qualitative look at the various solutions in Figure~\ref{fig:magnetic_overview}. We show here the 3D solutions, focusing on the magnetic field structure (the physical quantity common to all three models) in the inner corona up to $2.5 \, R_\odot$. We show results from the simulations without (top row) and with the far-side AR included (bottom row). The left panels show the PFSS results, the middle panels the WP results and the right panels the WP-AW results. Magnetic field lines are gray when closed, red when open with a positive polarity and blue when open with a negative polarity. The heliospheric current sheet (HCS) is also shown as a continuous black line around the Sun. {We also indicate whether a streamer is a helmet streamer contributing to the HCS (S) or a pseudo-streamer (P), and we attribute them numbers. We recall that helmet streamers separate coronal holes of opposite magnetic polarity, while pseudo-streamers overlie twin loop arcades and thus separate coronal holes of the same polarity \citep{Wang2007}.}
 
For the PFSS model (panels {a and d} of Figure~\ref{fig:magnetic_overview}), we observed that the HCS is rather flat and fully continuous, which means that open field lines are mostly located at the poles and closed field lines mostly near the equator. The strongest latitudinal modulation of the HCS is on the eastern limb, at least without the far-side AR (panel {a}). We observed one large helmet streamer on the western limb {(S1) surrounded by two pseudo-streamers (P1 and P2)} and one helmet streamer {(S2) with a pseudo-streamer in the southern hemisphere (P3) on the eastern limb}. 
With the inclusion of the {far-side AR} (panel {d}), the HCS becomes even flatter, even on the western limb. This is surprising since the AR is located closer to the eastern limb, so opposite to the western limb. With the far-side AR, the eastern streamers {appear} more complex, which is expected.

For the WP model (panels {b and e}), we observed that the {HCS has larger northward excursions from the equator near the eastern and western limbs. The increased complexity compared to the PFSS model is a result of the additional physics in the MHD model.}
For the case without the far-side AR (panel {b}), one can see {a streamer structure similar to the PFSS case (panel a) but with closed field lines that are more extended and more concentrated along the equator. We can also see that in this case, we have only one pseudo-streamer (P1) and two streamers (S1 and S2) on the eastern limb, while the pseudo-streamer on the western limb is absent.} 
When we include the far-side AR (panel {e}), the HCS is flattened but fragmented, with a large flux tube opening in the south on the eastern limb {(indicated by an arrow)}. The northern eastern streamers shift so that they are now entangled in the line of sight ({which creates the impression that S2 and S1 have merged}). 

Finally, for the WP-AW model (panels {c and f}), we observed that the results are similar to the WP model as they are both MHD models, {but even more complex due to the additional physics}. Without the far-side AR (panel {c}), the HCS is very modulated northward near both limbs, but remains continuous. On the western limb, {we have like in the PFSS case one helmet streamer (S4) and one pseudo-streamer (P1)}. On the eastern limb, {we have now three helmet streamers (S1, S2 and S3) with no pseudo-streamer}. With the inclusion of the far-side AR (bottom panel), the HCS also becomes more flattened but fragmented (with an even bigger southern flux tube opening than the WP model on the eastern limb, {indicated by an arrow}). The western streamers remain unaffected, while {the eastern streamers S2 and S3 seem to merge, while S1 becomes a pseudo-streamer (P2)}.

From this quick overview, we observed that all three models result in slightly different coronal structures, but that each of them is sensitive to the inclusion of the far-side AR. Surprisingly, including the AR does not only change the eastern limb (which is the closest one to the AR location), but also affects the western limb for all models. Changes on the western limb are crucial for space weather forecasting, as this is the side that is more likely to be magnetically connected to Earth due to the shape of the Parker spiral. This shows that ARs can have strong non-local effects on the global magnetic structure, and can thus modify the magnetic structure of the Sun at longitudes that are going to be impactful for Earth forecasting. In order to provide a more quantitative analysis, we next move on to other systematic comparisons between the three models.

\section{Impact of the far-side active region on closed magnetic structures}
\label{sec:results_wl}

\subsection{White-light emissions as a magnetic proxy}

In order to go beyond this quick overview, {we would need to use observations of the magnetic field structure of the corona for validation. However, there are no direct observations of the coronal magnetic field available. Instead, we can infer it, for example by using} white-light {(WL) polarized brightness} (pB) emissions images for comparison, as they are a {good} proxy of the magnetic structure of the corona. 
Such a proxy has already been used in previous studies, such as \cite{Wagner2022}, \cite{Perri2023} or \cite{Kuzma2023} in order to validate the number and location of streamers and pseudo-streamers.

{The WL pB solar emission is caused by Thomson scattering, that is, the scattering of photospheric radiation by free coronal electrons. This process is tied to the density of the corona, but the resulting images usually show the imprint of the magnetic field lines in the corona structure (with closed field lines generating more emissions than the open ones because they are denser).}

{We used data from two observatories. The first one is the COronal Solar Magnetism Observatory (COSMO) K-coronograph (K-Cor) from the High-Altitude Observatory (HAO) {\citep{DeWijn2012}}. K-Cor captures scattered white-light brightness images between 1.05 and 3 solar radii. We selected L1.5 data that are fully calibrated to $B/B_0$ in the 720-750 nm spectral range ($B_0$ being the mean solar brightness). These data were processed for instrumental effect and calibration as well as to remove sky polarization, corrected for sky transmission. The data are 10-minute averaged time series from February 15, 2021, at 18:01 UTC.\footnote{\url{https://doi.org/10.5065/d69g5jv8}} The images are $1024\times1024$ pixels in size.} 

{We also used data from the Large Angle and Spectrometric Coronagraph Experiment (LASCO) C2 coronagraph {\citep{Brueckner1995}} aboard the Solar Heliospheric Observatory (SoHO) {\citep{Domingo1995}}. LASCO/C2 captures scattered white-light brightness images between 2.2 and 6.5 solar radii. These pB data are in the spectral range 540-640 nm and were retrieved using the LASCO/C2 legacy archive \citep{Lamy2020} hosted by the MEDOC data center. These data were processed to remove instrumental effect and sky polarization, corrected for sky transmission, and calibrated to $10^{-10} B/B_0$ units. The data were taken at 15:01 UTC on February 15, 2021. The images are $512\times512$ pixels in size.}

{These images allowed us to capture the structure of the corona at a given time, but they can also be put together to give a more extended view of the temporal variations.} A proxy map developed by \cite{Poirier2021} (used already in \cite{Badman2022} and \cite{Perri2023}) uses LASCO/C2 data at around $3\;R_\odot$ to assemble them as a synoptic map over a Carrington rotation. This map gives an estimate of the streamer belt (SB), which can be assumed to host the HCS and act then as a proxy for it. It is however not a one to one comparison: the white-light brightness images capture the pseudo-streamers as well as the HCS streamer belt. Other effects can affect the detection, such as variations of brightness due to the orbital configuration between the Sun and Earth, or the passage of CMEs at the time of observation. The thickness is also not the same, as the HCS is usually thinner than the SB (for more details, see \citealt{Poirier2021}).

\begin{figure*}[htbp]
    \centering
    \includegraphics[width=\textwidth]{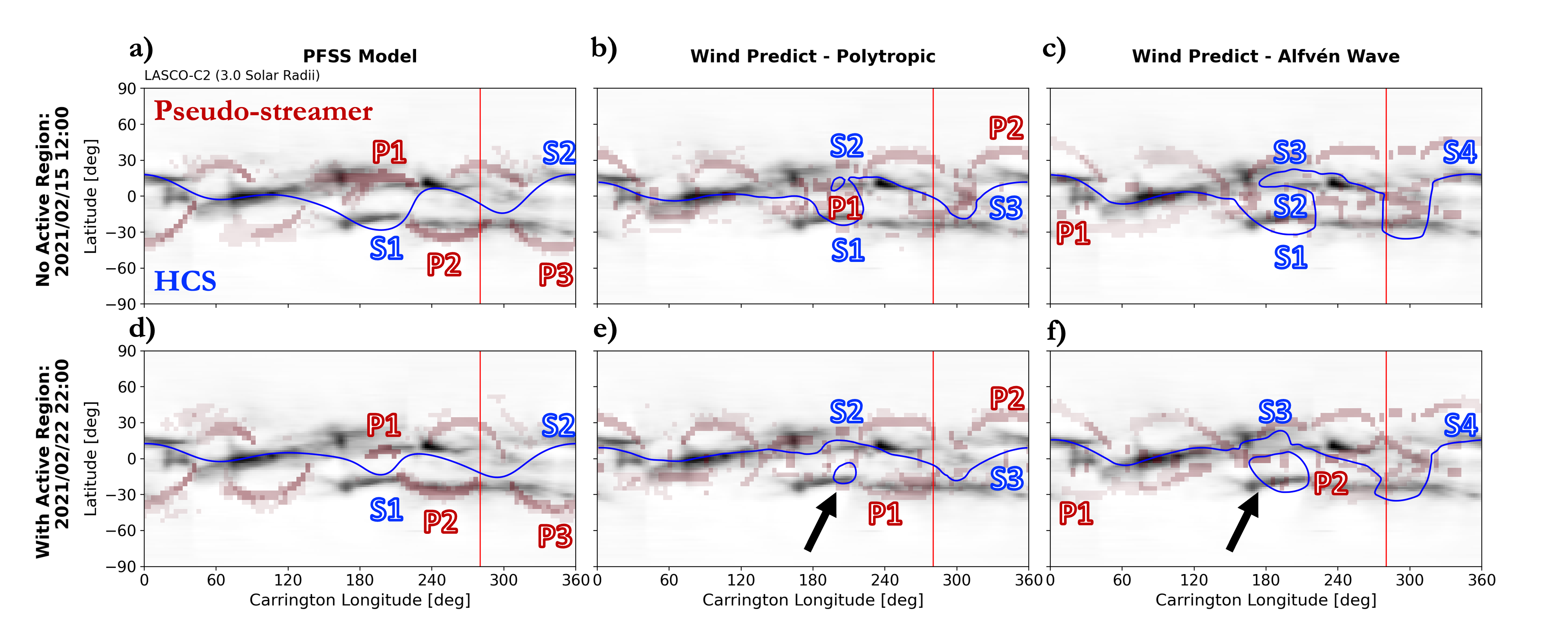}
    \caption{Comparison between white-light pB synoptic maps and streamers extracted from simulations for models without (top row) and with (bottom row) the far-side AR. The background is a synoptic map made using LASCO/C2 white-light brightness images over the corresponding Carrington rotation (CR2240). With the chosen color scale, the SB appears as dark regions. The HCS extracted from simulations is shown as a blue line. The pseudo-streamers extracted from simulations are shown as reddish regions. The longitude of the far-side AR on February 15, 2021, is shown with a vertical red line. {We keep the same nomenclature as in Figure~\ref{fig:magnetic_overview} for the identification of streamers and pseudo-streamers.}}
    \label{fig:streamers_all}
\end{figure*}

\subsection{Comparison with models}

In Figure~\ref{fig:streamers_all}, we show the comparison between white-light pB synoptic maps and streamers extracted from simulations for models without (top row) and with (bottom row) the far-side AR. The background is a synoptic map made using LASCO/C2 WL pB images over the corresponding Carrington rotation (CR2240). The HCS extracted from simulations is shown as a blue line {(delimiting the separation between positive and negative polarity in the $B_r$ field)}. The longitude of the far-side AR on February 15, 2021, is shown with a vertical red line. {We keep the same nomenclature as in Figure~\ref{fig:magnetic_overview} for the identification of streamers and pseudo-streamers.}

\begin{table*}[!h]
    \centering
    \begin{tabular}{c||c|c|c|c}
        WL analysis & HCS maximum deviation & HCS mean deviation & HCS standard deviation & Pseudo-streamers agreement \\ \hline \hline
        PFSS (no AR) & 47 degrees & 17 degrees & 14 degrees & 
        -- \\
        PFSS (AR) & 40 degrees & 15 degrees & 11 degrees & 
        - 28\%\\ \hline
        WP (no AR) & 20 degrees & 6 degrees & 5 degrees & 
        -- \\ 
        WP (AR) & 32 degrees & 7 degrees & 8 degrees & 
        - 0.9\%\\ \hline 
        WP-AW (No AR) & 34 degrees & 9 degrees & 8 degrees & 
        -- \\ 
        WP-AW (AR) & 43 degrees & 9 degrees & 10 degrees & 
        + 9\%\\ 
    \end{tabular}
    \caption{Summary of the changes observed in the closed magnetic field structures obtained by comparing with the WL images. We quantify for the three models (PFSS, WP and WP-AW) the changes in the HCS and the pseudo-streamers. For the HCS, we show the maximum, mean and standard deviation, while for the pseudo-streamers we show the variation of the agreement with the observations.
    }
    \label{tab:wl_numbers}
\end{table*}

With the chosen color map (which is an inverse gray map), the HCS is hosted in the darkest features exhibited. 
In order to validate our simulations, we would then need the {blue line} to go through all the darkest regions of the white-light emissions. However, if the blue line does not match the dark line, it can also mean that it is a pseudo-streamer instead of the HCS. 
This is why, in addition to the HCS, we also extract from all models a proxy for pseudo-streamers. The pseudo-streamer proxy was designed with the same idea as behind the squashing factor \citep{Titov2007}: here we compute the latitudinal displacement between neighboring footpoints of the magnetic field lines at 3 $R_\odot$, and if they find themselves separated by a sufficient amount above a certain threshold, then it probably means that they surround a pseudo-streamer structure. This allowed us to then estimate the location of pseudo-streamers, that we can then compare with the LASCO images. 
Using this proxy, pseudo-streamers are highlighted as pale red regions by selecting latitudinal displacements above $\sim 10$ degrees (darker shades of red then indicate larger displacements).
These additional red patches should then match the complementary dark regions that are not reached by the HCS. This is visible for example in panel {a}: the HCS matches the darkest features from longitudes 0 to 180, and then it oscillates between two dark branches north and south of the equator. With the pseudo-streamer proxy, we observed that the PFSS model considers that the northern branch is the HCS between longitudes 200 and 280, while the southern branch is actually matched by a pseudo-streamer to explain the dark region. It then reverses twice between longitudes 280 and 360, showing how difficult it is to distinguish between the HCS and pseudo-streamers. 
{We also want to add a word of caution regarding the data. The map is synoptic but not synchronic, and hence the further away we go in both directions from the vertical red line, the more outdated the data are going to be, explaining why pseudo-streamers around 60 degrees are going to differ significantly from the models.}
{We can then quantify the quality of the HCS and pseudo-streamers described in the models by comparison with the WL data. For the HCS, we can compute the path that goes through the darkest pixels of the map (which is the most likely to be the HCS path) and compute the deviation with the HCS extracted from the simulations.
For the pseudo-streamers, we can compute the percentage of covering between the ones extracted from the simulations and the ones from the simulations (using a simple threshold extraction to obtain the gray structures). Since this pseudo-streamer proxy is new and has not been fully quantified yet, here we focus more on the variations induced by the inclusion of the far-side AR than the absolute values. Results are summarized in Table~\ref{tab:wl_numbers}.}

We next describe the changes induced by the far-side AR for all models on the HCS and pseudo-streamers, and compare them with white-light observations for validation.
For the PFSS model ({panels a and d}), we observed that without the AR (panel {a}), the fit of the darkest regions by the HCS and pseudo-streamers is already good between longitudes 180 and 360. We see that the inclusion of the far-side AR (panel {d}) brings little changes to the structure of the HCS. It is already quite flat without the AR (maximum amplitude of 50 degrees), and becomes even flatter after emergence (maximum amplitude of 30 degrees). This is opposite to what one would expect since the inclusion of the AR should increase the complexity of the magnetic configuration. The HCS corresponds to the darkest regions of the white-light map, although the biggest region on the bottom right is barely reached. The region affected the most is between longitudes 120 and 240, which is actually eastward of the far-side AR location. 
{In Table~\ref{tab:wl_numbers}, we observed that the inclusion of the far-side AR tends to reduce all deviations to the HCS, which means that the general description is more accurate.}
Pseudo-streamers appear to match quite well the SB, although they have a more curved shape than the dark WL structures. When we include the far-side AR, this curvature is even accentuated in the vicinity of the AR (close to the vertical red line). 
{We observed that the inclusion of the far-side AR tends to decrease the quality of the description of the pseudo-streamers by 28\% due to this increase in their curvature.}

For the WP model ({panels b and e}), the structure of the HCS is more complex due to the MHD effects, which means that we have the plasma dynamics and the field structure is more physically consistent. Without the AR (panel {b}), we see more sharp modulations of the SB (maximum amplitude of 50 degrees). With the AR (panel {e}), the SB becomes more flattened (maximum amplitude of 25 degrees), with the southern dark region between longitudes 150 and 240 no longer being reached. Instead, an open tube opens in this region, disrupting the HCS. 
{In Table~\ref{tab:wl_numbers}, we observed that the inclusion of the far-side AR tends to slightly increase the mean and standard deviations, due to a significant increase in the maximum deviation.}
Pseudo-streamers are a bit more northward than for the PFSS, thus not matching perfectly with the SB. Including the far-side AR seems to reduce this discrepancy, but causes the same shift as seen before. 
{We observed that the inclusion of the far-side AR tends to barely affect the quality of the description of the pseudo-streamers (with a slight decrease of 0.9\%).}

For the WP-AW model {(panels c and f)}, we see similar effects. It is the model that best reproduces the path across the darkest regions. However, the inclusion of the far-side AR {(panel f)} tends to flatten it and remove the same region as for WP, which also turns into an open flux tube. 
{In Table~\ref{tab:wl_numbers}, we observed that the inclusion of the far-side AR tends to increase the maximum and standard deviation, but does not affect the mean deviation. We recall that the model is not fine-tuned, so these results could be improved with a dedicated parameter study (which is, however, not the object of this paper).}
Pseudo-streamers are even more northward and curved, but the inclusion of the far-side AR seems to reduce this discrepancy.
{We observed that the inclusion of the far-side AR tends to increase the quality of the description of the pseudo-streamers (with an increase of 9\% for the quality of the agreement). WP-AW is the only model that shows this significant improvement in pseudo-streamers, due to the more complex underlying physics.}

In conclusion, in all models the inclusion of the far-side AR tends to flatten the HCS. 
The PFSS is the model that reproduces best the pseudo-streamers, but without the far-side AR, as its inclusion introduces deviations. On the contrary, the WP-AW model performs best with the inclusion of the far-side AR. This shows the importance of the modelling behind the magnetic field structure.

\subsection{Synthetic emissions for Wind Predict with Alfvén waves}

For the PFSS and WP model, it is not possible to go beyond this proxy comparison with WL data because these models do not have a realistic enough thermodynamic structure to simulate WL emissions. {It was, however, possible for the WP-AW model to produce white-light pB synthetic images for quantitative comparison with the observations.} To do so, we used the TomograPy package from \cite{Barbey2013} as described in \cite{Parenti2022}. {This package uses as input the 3D density distribution from the global model to calculate the white-light coronal emission due to Thomson scattering. Then it integrates the result along the line of sight to simulate the detection by a specific instrument.} 

\begin{figure*}[htbp]
    \centering
    \includegraphics[width=\textwidth]{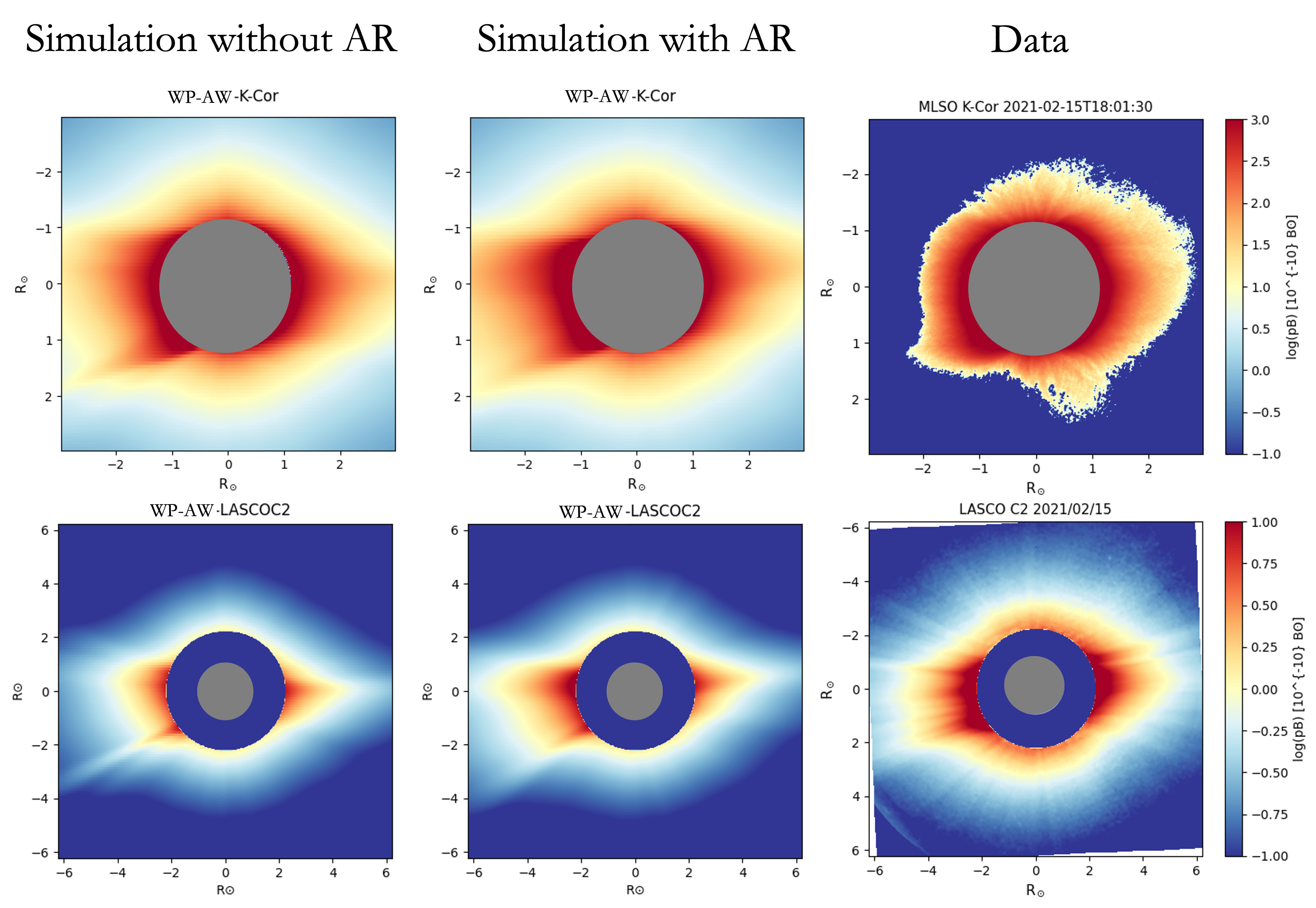}
    \caption{Comparison of white-light synthetic images with COSMO K-Cor and LASCO/C2 data of February 15 for the WP-AW model. The first line shows the results for COSMO K-Cor while the second line shows the results for LASCO/C2. The first column shows the results of the simulation for the input map of February 15 without the far-side AR, the second column shows the results of the simulation for the input map of February 22 with the inclusion of the far-side AR and the third column shows the real data for comparison. Each line has its own colorbar indicated to the right.
    }
    \label{fig:wl_2d}
\end{figure*}

First, we are going to comment on the magnetic structure inferred from the white-light images from the observations on February 15 (visible in Figure~\ref{fig:wl_2d} {in the right column}). On the western limb (on the right side of the image), the emissivity is dominated by a massive helmet streamer located near the equator. At least one pseudo-streamer could also be present slightly southward, but may not be on the plane perpendicular to the line of sight and is less intense than the main streamer. On the eastern limb (on the left side of the picture), we can clearly see in the LASCO data a three-fold structure. The K-Cor data are not so clear, but it seems reasonable to assume that the middle structure near the equator on the western limb is also a helmet streamer, while the northern and southern structures on the eastern limb are likely to be pseudo-streamers.

\begin{figure*}[htbp]
    \centering
    \includegraphics[width=\textwidth]{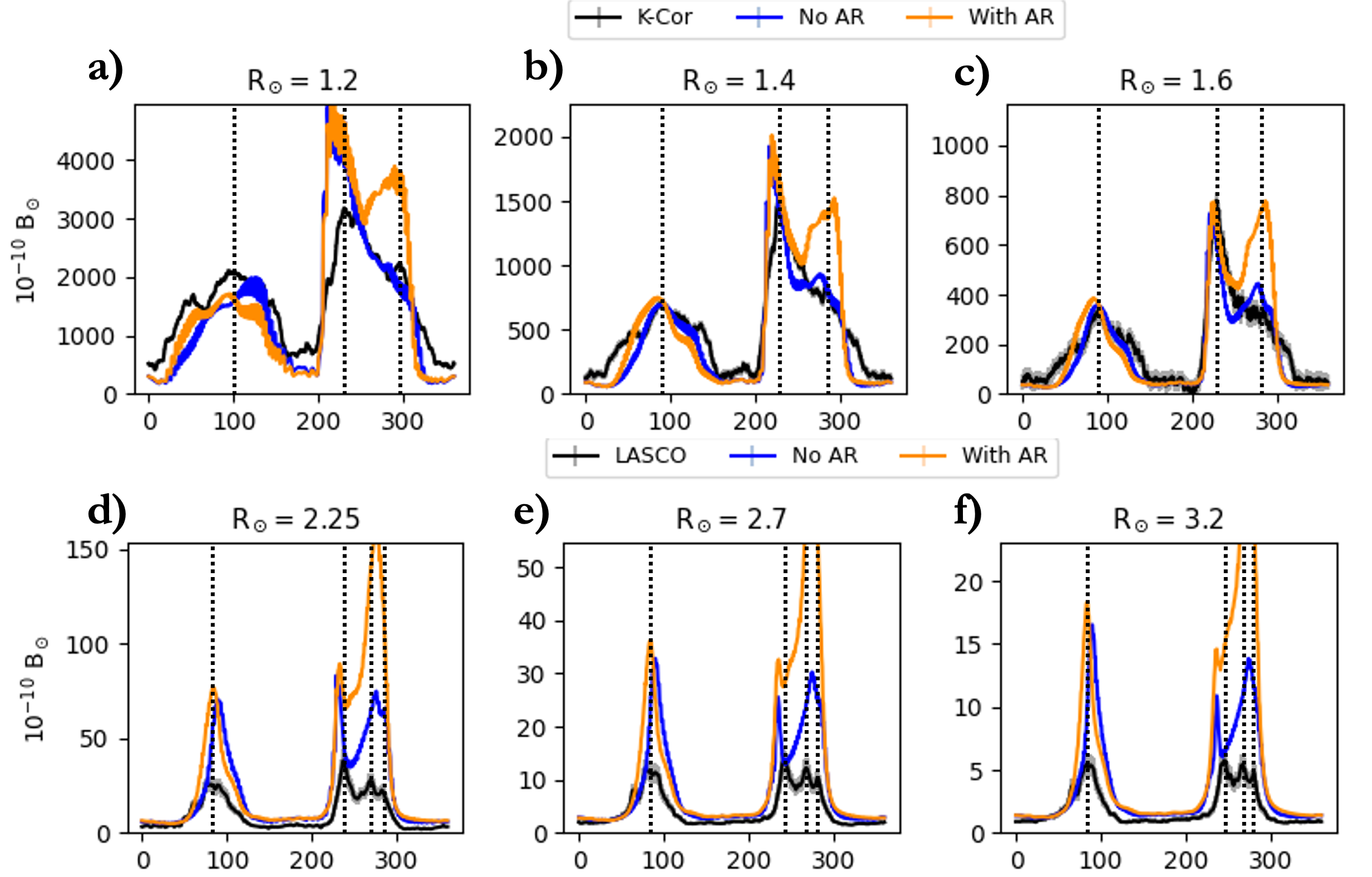}
    \caption{Comparison of 1D cuts of the synthetic white-light emission from the WP-AW simulations with data from COSMO K-Cor (top line) and SOHO/LASCO C2 (bottom line) from February 15. Each panel shows the comparison at respectively $1.2$, $1.4$, $1.6$, $2.25$, $2.7$ and $3.2$ solar radii. The simulation with WP-AW of February 15 without the far-side AR is shown in blue, while the simulation of February 22 with the far-side AR is shown in orange. Data from K-Cor and LASCO is shown with a black line. The gray areas show the error bars associated with the data. The x-axis shows the position angle in degrees, with 0 being the north pole of the Sun and the angle progressing in clock-wise direction (so that the western limb is shown from 0 to 180 degrees, and the eastern limb is shown from 180 to 360 degrees). {Streamer structures visible in the data are highlihgted using vertical dotted lines.}}
    \label{fig:wl_wp_aw_1d}
\end{figure*}

In order to quantitatively see the streamers in both observations and simulations, we perform latitudinal 1D cuts through the 2D images (both data and simulations), which we present in Figure \ref{fig:wl_wp_aw_1d}. Each panel shows the comparison at respectively $1.2$, $1.4$, $1.6$, $2.25$, $2.7$ and $3.2$ solar radii. The simulation with WP-AW of February 15 without the far-side AR is shown in blue, while the simulation of February 22 with the far-side AR is shown in orange. Data from K-Cor and LASCO is shown with a black line. {Streamer structures visible in the data are highlighted using vertical dotted lines.} The gray areas show the error bars associated with the data. 
{As explained in \cite{Parenti2022}, for K-Cor data the background noise in the pB is estimated at $3\,10^{-9} B/B_0$. For LASCO data, these error bars are estimated based on the work from \cite{Frazin2012}, where they are assumed to be 15\% in the whole field of view. More details can be found in \cite{Lamy2020}.} The x-axis shows the position angle in degrees, with 0 being the north pole of the Sun and the angle progressing in clock-wise direction (so that the western limb is shown from 0 to 180 degrees, and the eastern limb is shown from 180 to 360 degrees). 
As we said before, the K-Cor data {corresponds to structures close to the solar surface (between 1 and 3 solar radii}, so with the line-of-sight entanglement the three-fold structure of the western limb is less visible {(panels a, b and c, only two vertical dotted lines past 180 degrees).}
The large peak centered around 100 degrees is the large western streamer visible on the other limb. The three eastern streamers are more visible in the LASCO data {(three vertical dotted lines past 180 degrees in panels d, e and f)}, with three clear peaks at 230, 260 and 290 degrees at 3.2 $R_\odot$ {(panel f)}. 
For both cases (with and without the far-side AR in blue and {orange}) and both data (top and bottom panels), we observed that the western streamer is always present with a peak at the right angle around 100 degrees, which means there is a good agreement in alignment between the data and simulations. However, the intensity of the emission and the width of the streamer is only correct for K-Cor synthetic data in the top panel (this is not affected by the inclusion of the far-side AR). LASCO synthetic data in the bottom panel tend to overestimate the WL intensity, which is consistent with what was already found in \cite{Parenti2022}.
Regarding the eastern streamers, we observed clearly that for the synthetic K-Cor observations (top panel), the addition of the far-side AR does indeed generate an additional streamer peak around 300 degrees, that is clearly visible at 1.2 $R_\odot$ but also further away from the Sun. This could be due to an increase of the local density because of the inclusion of the far-side AR, which then leads to an increase of intensity in the white-light pB images. As for the LASCO synthetic data (bottom panel), it is less clear: The three streamers could be present at 2.25 $R_\odot$ for both simulations but then merge into two streamers further away from the Sun with also a tendency to overestimate the intensity (even stronger effect with the inclusion of the far-side AR).
In conclusion, we demonstrate further here that the inclusion of the far-side AR has an impact on the streamer structures, and that in particular it can improve their number and position when compared to observations. 
Next, we try to explain these changes by focusing on the open magnetic field distribution. 

\section{Impact of the far-side active region on open magnetic structures}
\label{sec:results_ch}

\subsection{Extreme ultra-violet maps as coronal holes proxy}

In order to know more precisely where the changes in streamer shapes come from, we examined the variations of open magnetic field lines in the lower atmosphere. We compared this structure with extreme ultra-violet (EUV) observations. This helped us identify the open field regions correlated with the darker patches seen in the EUV in the atmosphere, which are known as coronal holes \citep{Cranmer2009}.

{The EUV data comes from the 193\,\AA\ channel from the Atmospheric Imaging Assembly (AIA) instrument onboard the Solar Dynamics Observatory (SDO). They are L1 data, meaning that they have been corrected for instrumental effects and radiometrically calibrated (units of data number, or $Dn$). The images are $4096\times4096$ pixels in size.}

In particular, we used synoptic maps of EUV data to have the most consistent comparison with our models. 
These maps are constructed from full-disk images of the Sun from SDO/AIA. Each full-disk image is interpolated onto a Carrington longitude-latitude grid, 
then multiplied by a smoothly varying weight function, centered on the central meridian, and summed to produce a synoptic Carrington map. 
We use the official synoptic maps from SDO produced for each Carrington rotation.\footnote{\url{https://sdo.gsfc.nasa.gov/data/synoptic/}} The corresponding Carrington rotation for the studied period is CR2240, which started on January 22 and ended on February 18, 2021.

{The 193\,\AA\ band corresponds to the recommended wavelength to automatically extract coronal hole boundaries \citep{Wagner2022, Badman2022}. It is because it has a response function which peaks at the Fe XII temperature around 1.5MK, which makes it excellent to capture the contrast between magnetically open regions fuelling the fast solar wind, and hot, closed magnetic structures. 
The CHs from EUV data can then be automatically extracted using dedicated algorithms, such as the EZSEG algorithm developed by Predictive Science Inc.\ \citep{Caplan2016ApJ}.\footnote{The software is available as part of the EUV2CHM Matlab package from the Predictive Science Inc.\ website: \url{https://www.predsci.com/chd/}.} The details of the algorithm and its conversion to Python can be found in \cite{Perri2023}. We experimented with the optimal input parameters, and found that the best result was obtained with a connectivity of two neighbors, a first threshold at 20\,DN and a second threshold at 30\,DN (values for SDO/AIA 193\,\AA\ picture).}

\subsection{Comparison with models}

The WP-AW model produces a realistic solar atmosphere for which the EUV emission can be simulated. This allowed us to recover the coronal dimming. However the PFSS and WP models do not have a realistic enough solar atmosphere to model the EUV emission, and thus recover the coronal hole dimming. However, coronal holes also correspond to regions of open magnetic field lines, and this information can be retrieved from all models. We then proceed for the various models to find the boundaries between closed and open field lines at the {lower boundary of the domain}. We use a sphere of $400\times200$ seed points at $1.01\;R_\odot$. We followed the field lines to see if they reach the upper boundary of the computational domain {(see Table~\ref{tab:param_models} for the value for each model)}. If they do, they are defined as open field lines. If not, they are defined as closed field lines. This allowed us to retrieve contours of the open field line regions that we could directly compare with the coronal hole synoptic map. Similar comparisons have been performed in previous studies with positive results \citep{Badman2022}.

\begin{figure*}[htbp]
    \centering
    \includegraphics[width=\textwidth]{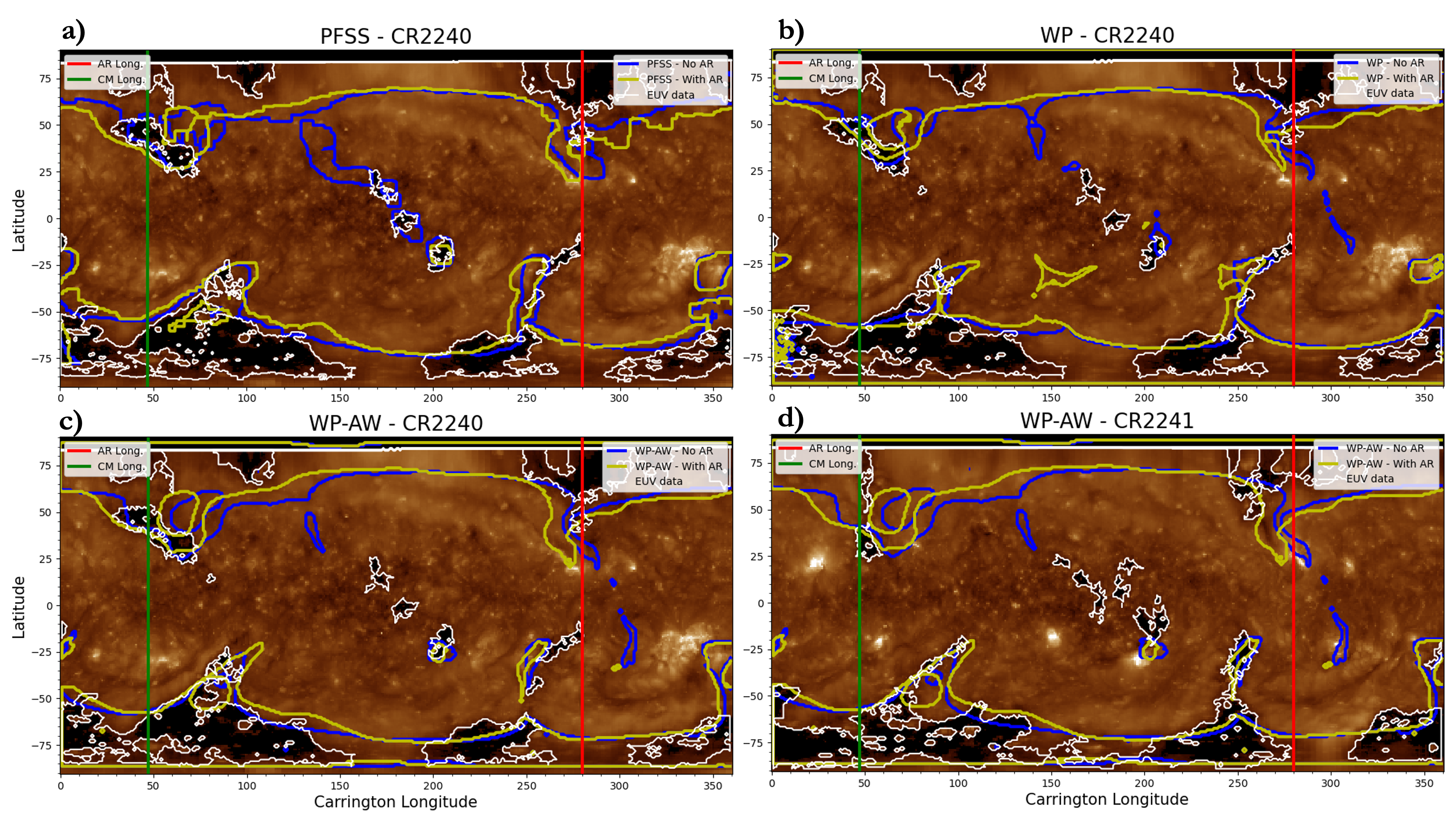}
    \caption{Comparison of the location of the coronal holes for the {PFSS} (top left panel), WP (top right panel), and WP-AW model (bottom left panel) with EUV observations by SDO/AIA 193 (background). We show the comparison with EUV data from Carrington Rotation 2240. Just for the WP-AW model, we also show the comparison with Carrington rotation 2241 (bottom right panel). Results extracted from observations are shown with white contours, while results obtained from numerical simulations are shown in color. The coronal holes corresponding to the simulation performed using the magnetic map from February 15 without the AR are shown in blue, while the ones using the magnetic map from February 22 with the AR are shown in yellow. We also indicate the longitude corresponding to the AR on the far-side with a red line, and the longitude corresponding to the central meridian as seen from the Earth with a green line.}
    \label{fig:ch_all}
\end{figure*}

We can thus plot in Figure~\ref{fig:ch_all} the comparison between CHs extracted from EUV observations, and CHs extracted from the open magnetic field lines regions {in the models}. 
Results extracted from observations are shown with white contours, while results obtained from numerical simulations are shown in color. The coronal holes corresponding to the simulation performed using the magnetic map from February 15 without the AR are shown in blue, while the ones using the magnetic map from February 22 with the AR are shown in yellow. We also indicate the longitude corresponding to the AR on the far-side with a red line, and the longitude corresponding to the central meridian as seen from the Earth with a green line.
We focus especially on the region between Carrington longitudes 210\textdegree\ and 330\textdegree, as it is the region where the far-side AR is located. 
We show for most panels the comparison with CR2240. However, since the emergence was on the far-side on longitude 280, the data was not yet updated on the asynchronous map, which means that the description of the CH on the right side of the map is 27 days late. We then also include the comparison of the WP-AW solution with CR2241 on the bottom right panel to show the difference.

We begin with the PFSS model (panel {a}). 
We observed that without the far-side AR (in blue), the PFSS model already provides a good description of the CH borders. It tends to overestimate the northern and southern coronal hole coverage by extending them to all longitudes, but this is the case for all models. We can also argue that this is the region where the CH from the EUV data are less reliable as well because of a lack of polar data, which can lead to projection effects and associated errors {(for more details, see Appendix~\ref{app:euv_mag})}. It also tends to overestimate the equatorial coronal holes. These are usually the most difficult to reproduce due to their smaller size, given the fact that we have considered the magnetic maps only up to $\ell_{max}=30$ in this study.
We can define a percentage of agreement for the definition of the CHs, which includes the number of pixels that are found both in observation and models to be inside or outside a CH divided by the total number of pixels.
We find that without the far-side AR, the percentage of agreement of the PFSS with the EUV data extraction is 73\%. 
Now with the inclusion of the far-side AR (in yellow), we observed slight changes in the CH shapes. The biggest change is probably concerning the equatorial CHs that almost disappear (except for the lower-right one). We can also notice that the fragmentation of the CHs is more realistic (especially visible between longitudes 50 and 100). 
With the inclusion of the far-side AR, the percentage of agreement of the PFSS with the observations remains similar at 72\%. There are however global changes, as the percentage of difference for the PFSS with and without the far-side AR is of 8\% of the pixels. This is probably the most surprising result here: all the CH borders are affected, not only the ones located close to the AR.

We now move on to the WP model (panel {b}).  
We recover similar properties compared to the PFSS model: northern and southern CHs are more extended than the data suggest, and equatorial CHs are more difficult to reproduce. Compared to the PFSS model, the CH structures are thinner.
We find that without the far-side AR, the percentage of agreement of WP with the EUV data extraction is 76\%. With the inclusion of the far-side AR, it remains similar at 74\%. There are however global changes, as the percentage of difference between WP results with and without the far-side AR is of 6\%. We observed a stronger shift of the CH borders up to 10 degrees in longitude, globally towards the center of the picture (longitude 180).

Finally, we analyze the results for the WP-AW model ({panels c and d}). Properties are similar to what is described before, but we observed that the matching for the CH borders is much better. For example, fragmentation for the southern CHs is already present without the inclusion of the far-side AR.
We find that without the far-side AR, the percentage of agreement of WP-AW with the EUV data extraction is 81\% {for panel c}. With the inclusion of the far-side AR, it remains similar at 80\%. There are however global changes, as the percentage of difference between WP with and without the far-side AR is of 7\%. The shift to the left close to the AR is even stronger (up to 20 degrees) and matches way better with CR2241.  

In conclusion, when we compare these results all together, we observe clearly that improving the physics of the model improves the quality of the CH description, with or without the far-side AR. Including the latter always changes CH borders by around 7\%. This change tends to lower by 1\% the quality of the CH description. However, what is interesting is that the change is global, even though the AR emergence is localized. This shows that coronal hole boundaries can evolve quickly even at minimum of activity, and that they are globally sensitive to local perturbations. It also explains what we saw earlier with the streamers: the inclusion of the far-side AR tends to reduce the amount of open magnetic field close to the equator, which leads to a global reconfiguration of the streamers.

\subsection{Synthetic EUV emissions for Wind Predict with Alfvén waves}


\begin{figure*}[htbp]
    \centering
    \includegraphics[width=\textwidth]{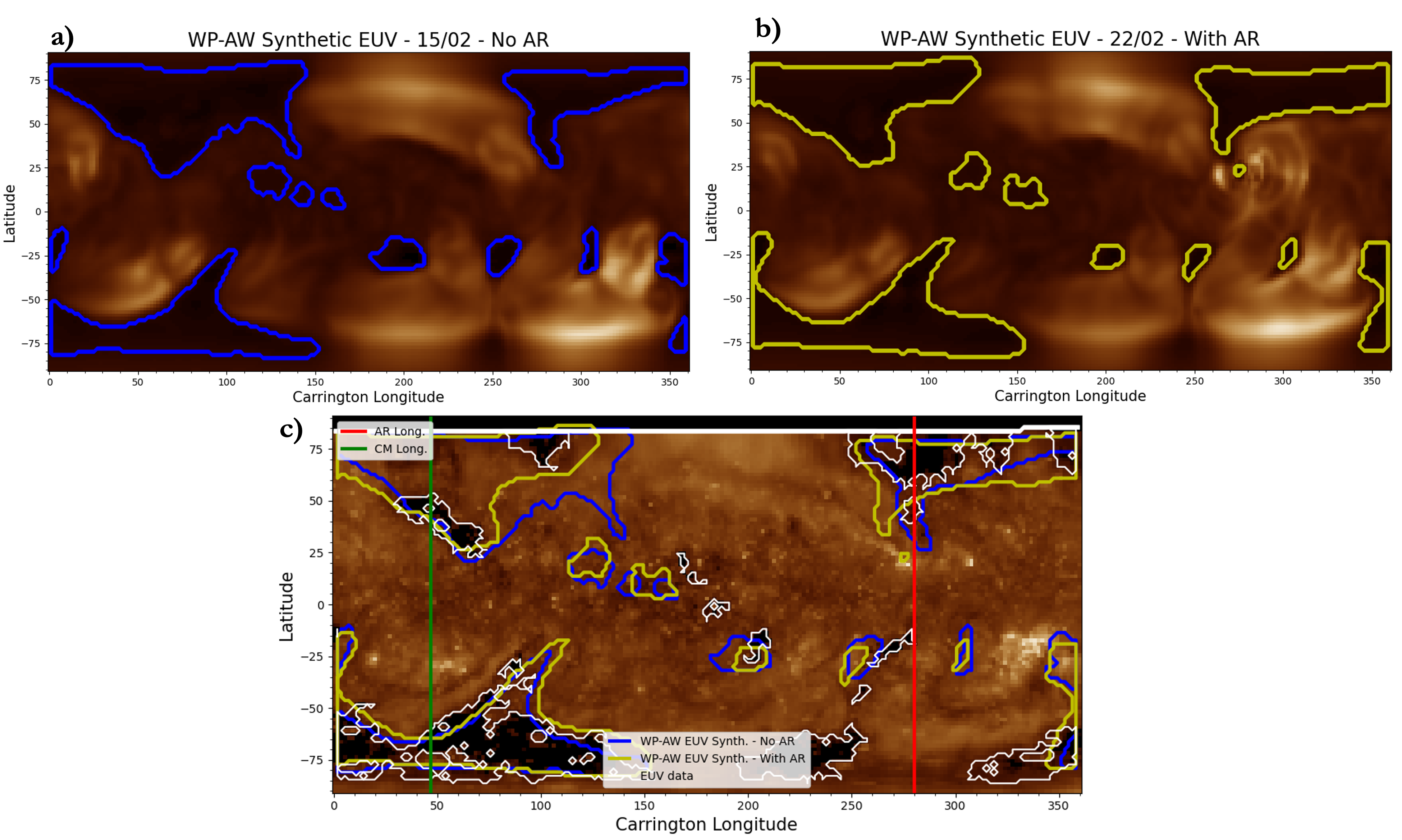}
    \caption{Comparison of synthetic EUV emission from WP-AW (top row) with EUV synoptic maps from SDO/AIA-193A (bottom row). On the top row, the left panel shows the results using the magnetic map from February 15 without the AR (blue contours), while the right panel shows the results using the magnetic map from February 22 with the AR (yellow contours). On the bottom row, we use the same design as described in the last Figure, where we compare contours from data (white) with WP-AW synthetic EUV contours without (blue) and with (yellow) the far-side AR. We also indicate the longitude corresponding to the AR on the far-side with a red line, and the longitude corresponding to the central meridian as seen from the Earth with a green line. The comparison is made with CR2240.}
    \label{fig:euv_comp}
\end{figure*}

The polytropic approximation we used for the coronal heating in WP did not allow us to use more refined techniques for comparison with EUV data since only the magnetic field is described realistically. Synthetic EUV emissions can, however, be generated from simulations to provide an accurate comparison \citep{Lionello2009}. Using the calibrated response of a given EUV instrument \citep[e.g., SDO/AIA, see][]{Landi2013} as a function of the number density and the temperature of the plasma, the image is produced integrating the volume response along the line of sight. Wind Predict-AW has been shown to agree well with AIA 193 A images during solar minimum phases \citep{Parenti2022}. To create synoptic maps, we use here the central meridian response and perform a rotation of the simulation to account for all Carrington longitudes. To extract the contours, we apply the same EZSEG algorithm as with the EUV data, where we just adjusted the parameters to fit the synthetic emission levels (first threshold at 14, second threshold at 16).

The results of this comparison can be seen in Figure~\ref{fig:euv_comp}. The top row shows synthetic EUV emission from WP-AW, while the bottom row shows the comparison directly with the EUV synoptic maps from SDO/AIA-193 for CR2240. For the top row, the left panel shows the results using the magnetic map from February 15 without the AR (blue contours), while the right panel shows the results using the magnetic map from February 22 with the AR (yellow contours). 
We observed that the resolution is not the same as the previous results with the open magnetic field. It was slightly downgraded for reasons related to computing resources for the generation of the EUV synthetic maps, so we interpolated the EUV data map to match it so that the comparison is relevant. 
At first glance, we observed that the emergence of the far-side AR is very visible in the synthetic EUV {(panel b)}, with a clear enhancement of luminosity at the longitude 280. The structure however appears to reflect multiple emergence and not only one, which could be due to the spherical harmonics filtering applied, or additional fainter surrounding magnetic structures.
What is interesting in this other approach is that we capture the north and south limits of the CHs better. Instead of being extended to all longitudes, northern and southern CHs are fragmented with limits that match the observations. This shows that EUV emissions are not a one-to-one proxy for the magnetic field open regions, which is something we can prove only with simulations {(for more details regarding this discussion, see Appendix~\ref{app:euv_mag})}. Regarding the far-side inclusion, we reached the same conclusions as with the open magnetic field, with less equatorial CHs and shifting of northern CHs close to the AR.

\section{Impact of the far-side active region for heliospheric connectivity}
\label{sec:results_insitu}

\subsection{In situ magnetic measurements}

In this section, we focus on large-scale magnetic structures that shape the interplanetary medium and are crucial to space weather forecasting, as they have an impact on the propagation of transients. Typically, such large structures are more difficult to get with the PFSS model, and thus it is extremely important to extract them from MHD models. These structures usually require an additional current sheet model (such as the Schatten one; see for example \citealt{Pomoell2018}), and even with it, it is usually challenging to compute in situ magnetic field data far away from the Sun due to numerical reconnection. For these kinds of structures, global observations are also more difficult to get for comparison. Here, we focus on in situ observations in order to validate and discuss the large-scale magnetic structures.


\begin{figure*}[htbp]
    \centering
    \includegraphics[width=\textwidth]{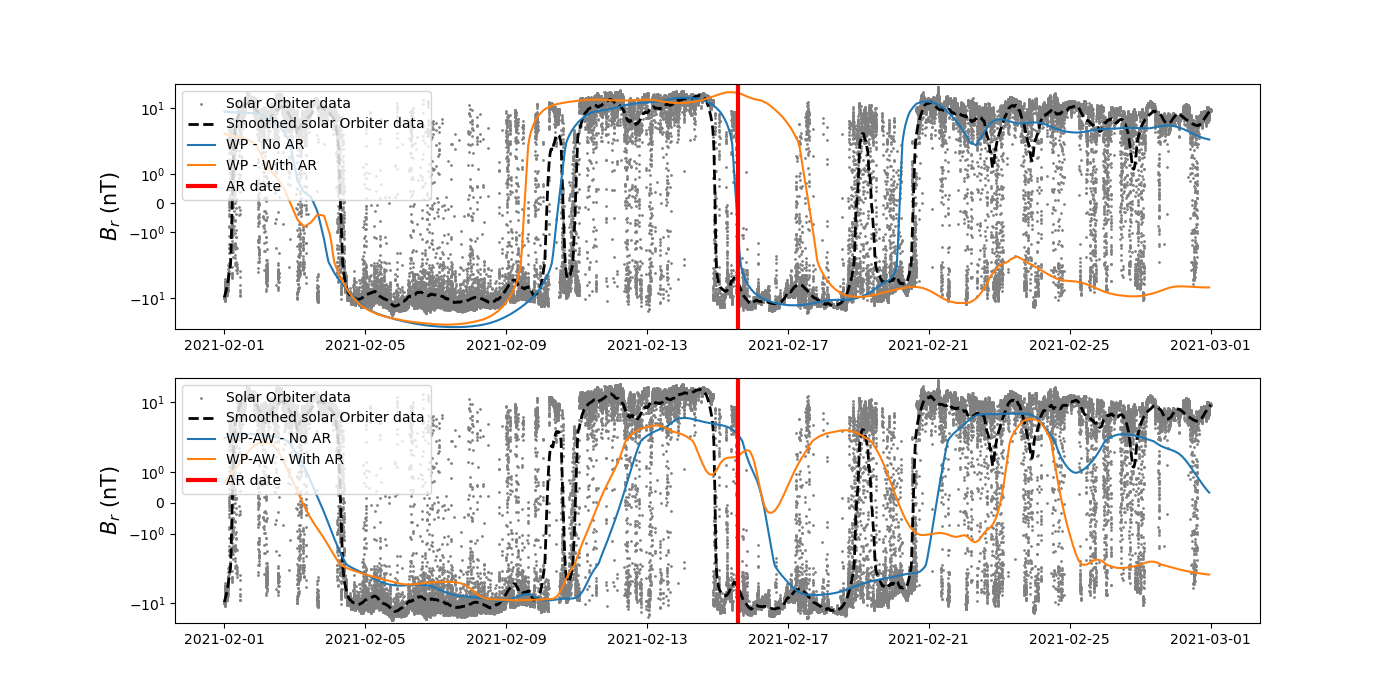}
    \caption{Comparison between Solar Orbiter/MAG measured radial field and predictions for the WP model (top panel) and the WP-AW model (bottom panel). The magnetic field corresponding to the simulation performed using the magnetic map from February 15 without the far-side AR is shown in blue, while the one using the magnetic map from February 22 with the far-side AR is shown in orange. For context, we show the data over the full month of February 2021 (in gray), as well as their smoothing (in black dashed line). We also indicate the date of the emergence of the far-side AR with a vertical red line.}
    \label{fig:in_situ}
\end{figure*}

To do so, we assess the in situ properties of the magnetic field along Solar Orbiter trajectory which we plot in Figure~\ref{fig:in_situ} as measured by the MAG instrument \citep{Horbury2020}, as it was passing on the far-side over the month of February 2021 (in gray). We use the 1-minute cadence L2 data {downloaded from the AMDA database \citep{Genot2021}, with a quality flag oscillating between two and three over this period}. It is clear however that numerical simulations will have difficulties reproducing this temporal resolution. This is why we also show the smoothed data (in black dashed line), obtained by applying a Savitzky-Golay filter with a window size of 1501 points and a 3rd-order polynomial \citep{Savitzky1964}. For this configuration, positive radial magnetic field means the spacecraft is connected to northern coronal holes, while negative radial magnetic field means it is connected to southern coronal holes.  

\subsection{Comparison with models}

To extract similar information from the numerical simulations, we use a virtual spacecraft that has the same trajectory in the corresponding frame of the simulation, and we plot the radial magnetic field at its location. For both models, we use a ballistic projection following the Parker spiral, extrapolating from $20 \, R_\odot$ the magnetic field assuming a 400\,km/s solar wind speed \citep{Neugebauer1998}. This value has been selected because it is the most probable in the equatorial plane during a period close to minimum of activity. {We could have used the actual measurements of the solar wind at Solar Orbiter for a more realistic result. However, SWA data are not available during this time period (which corresponds to the cruise phase of Solar Orbiter). Our simulations do not include the orbit of Solar Orbiter into their numerical domains (see Table~\ref{tab:param_models}), so instead we have estimated the solar wind speed at Solar Orbiter using the fast HUXt model \citep{Riley2011, Barnard2022}. During this period, we find indeed a solar wind speed around 400\,km/s on average.} The WP model tends to have more numerical diffusion than WP-AW, and hence is more prone to numerical reconnections that will decrease the magnetic field amplitude for non-physical reasons. This means that we need to apply a multiplying factor of 20 to get similar amplitudes to make better comparison with the data. We recall that here we are more focused on the polarity of the field than its actual values, and that the multiplying factor does not affect this property.

We compare results for the WP model (top panel) and the WP-AW model (bottom panel) in Figure~\ref{fig:in_situ}. The magnetic field corresponding to the simulation performed using the magnetic map from February 15 without the far-side AR is shown in blue, while the one using the magnetic map from February 22 with the far-side AR is shown in orange. 
We also indicate the date of the emergence of the far-side AR with a vertical red line. We see that, due to the lack of resolution, both simulations have less variations of the polarity than the real data, as expected, but do match the smoothed data. For the WP model, the main differences are close to the date of the AR emergence: the positive polarity lasts until February 17, and thus matches better the observations. It is also the case around February 9, where the positive polarity is caught earlier on. However, the positive polarity from February 21 is not captured anymore (but this period is less critical, as it is far from the AR date and thus the period of interest). For the WP-AW model, we see similar effects, but stronger: the positive polarity patch from February 17 to 21 which is completely missed without the AR is suddenly {well} captured with the inclusion of the far-side AR. This means that the inclusion of far-side data could improve both forecasting and hindcasting of the connectivity to the spacecraft. However the effect of a single AR with the current resolution of MHD codes is not clear, so confirmation with better resolution or with a more active far-side would be needed in future work. We expect this effect to be amplified when considering higher-resolution simulations, that will be able to capture smaller and smaller scales and that are left for future work.

{We summarize all the results found so far from all previous sections in Table~\ref{tab:results}. We move to the discussion section in order to attempt to explain all these results.}

\begin{table*}[]
\centering
\begin{tabularx}{\linewidth}{c | X}
  \toprule
  \textbf{Analysis} & 
    \hfill \textbf{Main results from the far-side AR inclusion} 
    \hfill\null \\
  \midrule
  \makecell{\textbf{Map analysis} \\ Unsigned flux increase} & 
    \hfill \makecell{\textbf{Original map} \\ 8\% of total map flux}
    \hfill \makecell{\textbf{Filtered map} \\ 4\% of total map flux}
    \hfill\null \\
  \midrule
  \makecell{\textbf{Model analysis} \\ 3D overview \\ \hspace{1cm} \\ HCS \\ \hspace{1cm} \\ WL emissions \\ \hspace{1cm} \\ Coronal holes \\ EUV emissions \\ Connectivity \\ to spacecraft} & 
    \hfill \makecell{\textbf{PFSS} \\ Concentration of eastern limb \\ streamers towards the equator \\ Flattening of HCS \\ \hspace{1cm} \\ N/A \\ \hspace{1cm} \\ From 73\% to 72\% agreement \\ N/A \\ N/A \\ \hspace{1cm}} 
    \hfill \makecell{\textbf{Wind Predict} \\ Merging of eastern \\ limb streamers \\ Flattening \\ and breaking of HCS \\ N/A \\ \hspace{1cm} \\ From 76\% to 74\% agreement \\ N/A \\ Improvement of positive \\ polarity connection} 
    \hfill \makecell{\textbf{Wind Predict-AW} \\ Radial extension \\ of eastern limb streamers \\ Flattening \\ and breaking of HCS \\ Improvement of southern \\ pseudo-streamer emission \\ From 81\% to 80\% agreement \\ Improvement of northern CH \\ Improvement of positive \\ polarity connection} 
    \null \\
  \bottomrule
\end{tabularx}
\caption{{Table summarizing the main results for each section and models from the study.
}}
\label{tab:results}
\end{table*}

\section{Discussion}
\label{sec:discussion}

We want to discuss here some of the results found above {summarized in Table~\ref{tab:results}}. In particular, we want to bring forward some elements that may explain why the local impact of this particular far-side AR is not necessarily as strong as anticipated, and also why the streamer structure is so sensitive for most models.

\subsection{Magnetic precursor to the active region}

\begin{figure*}[htbp]
    \centering
    \includegraphics[width=\textwidth]{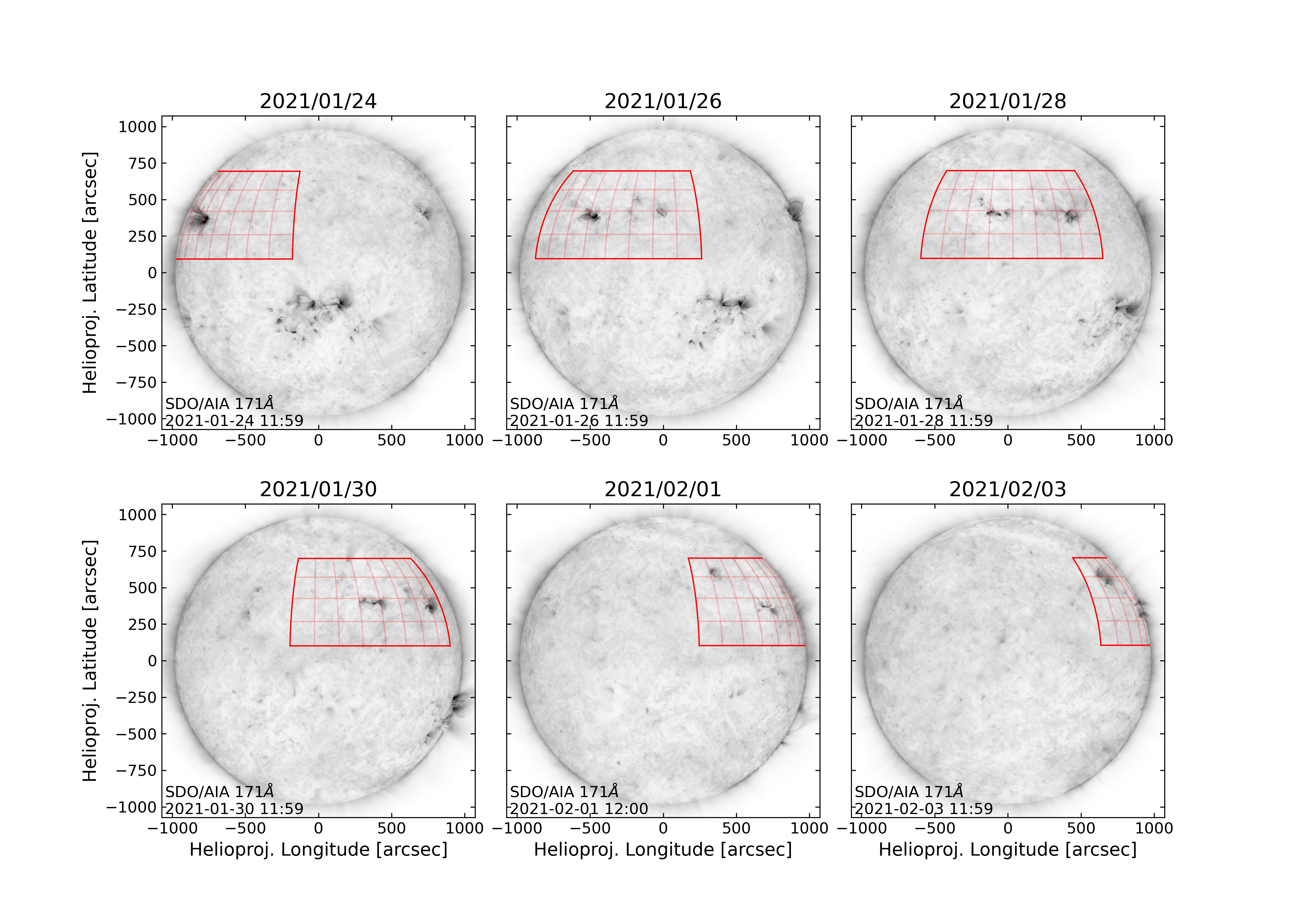}
    \caption{SDO/AIA 171 observations of the AR precursor from January 26 to February 3, 2021. We show for every two days the full-disk view of the Sun, with the region of interest marked with a red grid. {The grayscale makes intense magnetic structures such as ARs appear black, so that one can see a magnetic precursor at the same latitude as the AR examined in this study.}}
    \label{fig:ar_tracking}
\end{figure*}

One obvious reason for the not so strong local impact of the far-side AR could of course be its small flux, which we quantified to be only twice the surrounding quiet Sun flux in section \ref{sec:methodology}, and only 4 to 8\% of the total unsigned flux of the Sun. Some AR can increase the local unsigned flux by a factor of at least 10, which of course would be expected to generate more impact \citep{Derosa2018}. Another explanation as to why the flux increase is so low could be that the surrounding magnetic field is actually not so quiet due to previous emergence. We can check this theory by monitoring the AR location before its emergence. In Figure~\ref{fig:ar_tracking}, we observed the region where the far-side AR emerged when it was facing the Earth-side during the previous rotation, more specifically from January 26 to February 3, 2021 (which is two weeks prior to the AR emergence). We used SDO/AIA 171\,\AA\ data to track it, with a red grid marking the region of interest. We show for every two days the full-disk view of the Sun to put the region in context. This shows clearly that when the region of interest was facing Earth, there was a decaying AR which was present. This previous AR was labeled AR12800, while the new one that emerged on the far-side that we studied was labeled AR12803. When monitoring Solar Orbiter far-side data, we observed that AR12800 faded between February 3 and 7 and that AR12803 emerged at a similar location on February 15. This certainly explains why the impact of the AR was reduced. Not only is the AR of medium intensity for the solar standards, but it also emerged at the same location as a magnetic precursor  visible one week before. This shows the necessity of putting ARs in perspective in time; the intensity of the AR is not the only component that determines its impact, as the global solar context before its emergence also has an influence.

\subsection{Interpretation using the source surface radius}

\begin{figure*}[htbp]
    \centering
    \includegraphics[width=\textwidth]{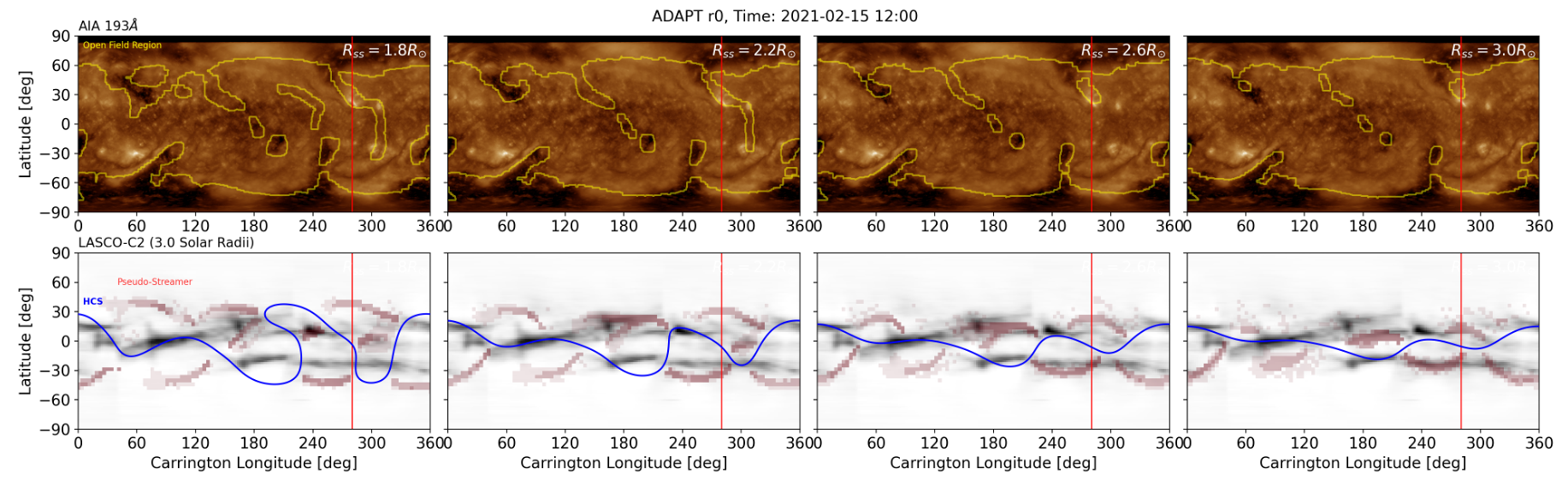}
    \caption{Study of the influence of the source surface radius ($R_{ss}$) for the PFSS model. In the top row, we show EUV synoptic maps made from SDO/AIA 193 observations, where we superimpose the boundaries of open magnetic field regions extracted from the PFSS model with yellow contours. In the bottom row, we show the comparison between the LASCO/C2 white-light synoptic maps at 3 $R_\odot$ with the PFSS model using as input the ADAPT map of February 22, 2021, at 22:00UT (with the AR). From the PFSS, we extract the HCS (blue line) as well as a proxy for pseudo-streamers (red patches). The different panels show this comparison for various $R_{ss}$ values (from left to right: 1.8$R_\odot$, 2.2$R_\odot$, 2.6$R_\odot$ and 3.0$R_\odot$). We indicate the longitude of the AR at the date with a vertical red line.}
    \label{fig:streamer_rss}
\end{figure*}

{How can we explain the global impact of the AR?}
With the PFSS model, we can perform simple tests where we vary the height of the source surface, which in return changes the limit between open and closed magnetic field lines, as well as the flux tube expansion rate. The result of this study is shown in Figure \ref{fig:streamer_rss}. In the top row, we show EUV synoptic maps made from SDO/AIA 193 observations, where we superimpose the boundaries of open magnetic field regions extracted from the PFSS model with yellow contours. In the bottom row, we show the comparison between the LASCO/C2 white-light synoptic maps at 3 $R_\odot$ with the PFSS model using as input the ADAPT map of February 22, 2021, at 22:00UT (with the AR). From the PFSS, we extract the HCS (blue line) as well as a proxy for pseudo-streamers (red patches). The different panels show this comparison for various $R_{ss}$ values (from left to right: 1.8$R_\odot$, 2.2$R_\odot$, 2.6$R_\odot$ and 3.0$R_\odot$). We indicate the longitude of the AR at the date with a vertical red line.

We can clearly see in Figure \ref{fig:streamer_rss} that by changing the height of the source surface, we change the shape of the HCS which looks more similar to what is obtained with the MHD models without the AR. There is also a critical value where we transition from two streamers to three streamers at the longitude of the AR. For 2.6 $R_\odot$ (which is close to the value used of 2.5 used in the previous study), the line crosses the two pseudo-streamers, while for 1.8, the shift of the pseudo-streamer makes the crossing unclear. A similar effect can be observed with the PFSS model using the map without the AR as input. Similarly, when we look at the CH structure evolution with the PFSS, we find that for smaller $R_{ss}$, we have bigger CHs, while for larger $R_{ss}$, equatorial CHs close and polar CH boundaries shift, just like with the inclusion of the far-side AR. 
This analysis shows that the PFSS case without the AR seem to be equivalent to a PFSS model with a lower source surface (which means that magnetic field lines open closer to the solar surface), and that the inclusion of the far-side AR has a similar effect as to increasing the source surface height globally. This could explain most of the effects described in the previous sections. With a lower source surface, it is easier to open coronal holes (hence the equatorial CHs better reproduced) and to reproduce pseudo-streamer structures. This allowed for a more complex HCS, but this does not necessarily equate to a better connectivity estimation. All the previous results can therefore be explained by the fact that the inclusion of the far-side AR not only increases locally the magnetic flux, but also globally increases the height of closed field loops, which explains the global effects observed. Including missing far-side flux without modifying running parameters to mitigate this effect might then yield results that do not match better observations, especially for models that are not fully self-consistent contrary to the WP-AW model. This discussion opens a reflection on the necessity to adapt our typical run parameters to the input of magnetic flux, which could be determined by exploratory parametric studies (which is of course outside the scope of this paper).

\section{Conclusion}
\label{sec:conclusion}

%

In this work, we have estimated the impact of far-side magnetic field structures observed by Solar Orbiter on 3D coronal and heliospheric wind simulations. We analyzed the case of a far-side emerging AR on February 15, 2021, causing a local increase of magnetic flux by a factor two, which corresponds to 8\% of the total flux of the input map (4\% of the processed map). Since synchronic magnetic maps using Solar Orbiter/PHI data are not yet available, we made a proxy experiment where we used two GONG-ADAPT magnetic maps. The first one was made on February 15 and does not include the far-side AR since it was not visible from Earth, while the second one was made on February 22 and includes the far-side AR once it became visible on the Earth-facing disk. We then used the magnetic maps to initialize global coronal numerical simulations, with a filtering by spherical harmonic modes limited to $\ell_{max}=30$ (25 for WP-AW). We used three different types of models: a PFSS semi-empirical extrapolation, a polytropic 3D MHD WP model based on the PLUTO code, and finally its upgrade with realistic description of the transition region and the coronal heating using Alfvén waves (i.e., WP-AW). For each case, we assessed the effect of the inclusion of the far-side AR using comparisons with both remote-sensing observations, such as SDO/AIA EUV maps or COSMO/K-Cor and SOHO/LASCO C2 white-light pB images, and in situ diagnosis, such as from Solar Orbiter/MAG measurements. We compared all three models between each other to assess the impact of the underlying physical hypothesis on the accurate representation of the global corona and offered some physical insights as to why these effects are observed. We summarize our main conclusions with the following points.

Adopting a source surface of 2.5 $R_\odot$ (canonical value), the PFSS model manages to reproduce quite well the distribution of coronal holes as well as the HCS positioning, which is expected for this kind of low-activity configuration \citep{Kruse2021}. To be more specific, the CH distribution is correct at 73\% without the AR and at 72\% with the AR. The HCS has a reduction of latitudinal amplitude of around 20 degrees, which makes it globally more flat. The streamer distribution, however, appears to be better reproduced without the inclusion of the far-side AR because it causes changes in the streamers' shape and position.

Both MHD models perform well for the distribution of coronal holes, with similar results. To be more precise, the WP CH distribution is correct at 76\% without the AR and at 74\% with the AR, while the WP-AW CH distribution is correct at 81\% without the AR and at 80\% with the AR, which means both models provide a better CH description than the PFSS.
However, when it comes to the streamers, the WP model does not seem to be improved by the inclusion of the far-side AR, while the WP-AW model clearly is.
This can be demonstrated without ambiguity since the WP-AW model allowed us to produce synthetic EUV and white-light emission from which we could make more quantitative comparisons with the observations. The WP-AW model tends to overestimate the white-light intensity but manages to recover the streamer structure with the inclusion of the far-side AR. The EUV synthetic emissions also better recover the borders of the EUV CHs without the longitudinal extent. 
Both models also allow for in situ diagnosis, which is not possible with the PFSS model. Both models exhibit a more flattened HCS with the inclusion of the AR, just like the PFSS. They also show switches in connectivity, with the positive polarity between February 15 and 20 being better captured with the inclusion of the far-side AR. 

Concerning the overall impact of the inclusion of the far-side AR, it does not have a strong local impact, as coronal hole boundaries and streamer shapes are only slightly affected. This may be due to the fact that this specific region of interest was not very intense (flux only 8\% of the total surface flux) and also hosted the seat of a precursor AR two weeks before, whose influence may still be felt. The strongest effect is actually non-local. The inclusion of a single far-side AR of medium magnetic intensity drives big changes to the global magnetic structure of the {Sun}. Coronal {hole} boundaries on the Earth-facing side are affected as well, as the HCS shape may differ, and the comprehension of the in situ connectivity changes. Our understanding of this global change is that the inclusion of the far-side AR is equivalent to an enhanced source surface with closed loops reaching higher layers of the corona, which in turn closes the equatorial CHs and therefore affects the streamers' locations, the HCS, and the resulting connectivity. This means that without information from the far-side, simulations of even the Earth-facing corona may be biased, which shows the need to take them into account. However, including the far-side AR without modifying the parameters of the simulation works only for the most realistic models, such as WP-AW, and may require more exploratory studies to improve forecasts for faster models. 

This work attempts to prepare the community for the transition from asynchronous synoptic maps as input to global coronal models to fully synchronic maps enabled by the Solar Orbiter mission. We performed a preliminary study in anticipation of this transition. A full test will only be possible when such maps are completed, which hopefully should be in the near future, for Solar Orbiter data at least \citep{Sinjan2023A&A}. This study experimented with a single far-side AR. Therefore, a more systematic analysis would be required in order to validate our results. We have also made the choice to study only one AR during quiet solar conditions. During periods of higher solar activity, the photospheric magnetic field will evolve more rapidly due to increased rates of flux emergence such that far-side observations become more important. Finally, this study has been made with three coronal models, and although they do cover a wide range of categories of models (PFSS, MHD, MHD with realistic heating functions), it would be interesting to see if such results hold for other models as well. Our future work will focus on more Solar Orbiter events as well as the development of even more diagnosis tools in order to further increase our capacity to compare observations and models. {We hope that this kind of systematic comparison can also help improve the current models in both finding the best input parameters and pointing out the most interesting physics to include. A very natural follow-up work will be to use these results to find the best parameter space for WP-AW with and without the inclusion of far-side data in order to provide the best forecasting as well as the best understanding of solar and heliospheric physics.}

\begin{acknowledgements}
    This work was supported by the CNRS and INSU/PNST program, CNES Solar Orbiter and “Météorologie de l’espace” funds, the ERC Synergy grant WholeSun No.810218, and the ANR STORMGENESIS \#ANR-22-CE31-0013-01.
    Computations were carried out using CNRS IDRIS facility within the GENCI A0130410293 and A010810133 allocations and a local mesocomputer founded by DIMACAV+.
    The authors are grateful to A. Mignone and the PLUTO development team.
    Courtesy of the Mauna Loa Solar Observatory, operated by  the High Altitude Observatory, as part of the National Center  for Atmospheric Research (NCAR). NCAR is supported by the National Science Foundation. 
    This work makes use of the LASCO-C2 legacy archive data produced by the LASCO-C2 team at the Laboratoire  d’Astrophysique de Marseille and the Laboratoire Atmosphères, Milieux, Observations Spatiales, both funded by the  Centre National d’Études Spatiales (CNES). LASCO was built  by a consortium of the Naval Research Laboratory, USA, the  Laboratoire d’Astrophysique de Marseille (formerly Laboratoire d’Astronomie Spatiale), France, the Max-Planck-Institut für Sonnensystemforschung (formerly Max Planck Institute für Aeronomie), Germany, and the School of Physics and  Astronomy, University of Birmingham, UK. 
    SOHO is a project of international cooperation between ESA and NASA. 
    This work used data provided by the MEDOC data and operations centre (CNES / CNRS / Univ. Paris-Saclay), \url{http://medoc.ias.u-psud.fr/}.
    Courtesy of NASA/SDO and the AIA, EVE, and HMI science teams. 
    This work utilizes data produced collaboratively between AFRL/ADAPT and NSO/NISP.
    Solar Orbiter is a mission of international cooperation between ESA and NASA, operated by ESA.
    The EUI instrument was built by CSL, IAS, MPS, MSSL/UCL, PMOD/WRC, ROB, LCF/IO with funding from the Belgian Federal Science Policy Office (BELSPO/PRODEX PEA 4000112292); the Centre National d’Etudes Spatiales (CNES); the UK Space Agency (UKSA); the Bundesministerium für Wirtschaft und Energie (BMWi) through the Deutsches Zentrum für Luft- und Raumfahrt (DLR); and the Swiss Space Office (SSO).
    {Solar Orbiter magnetometer data was provided by Imperial College London and supported by the UK Space Agency.}
    {Data analysis was performed with the AMDA science analysis system provided by the Centre de Données de la Physique des Plasmas (CDPP) supported by CNRS, CNES, Observatoire de Paris and Université Paul Sabatier, Toulouse.}
\end{acknowledgements}

%
%

\bibliography{solo2}
\bibliographystyle{aa}

\appendix

\section{Comparison between coronal hole boundaries derived from open magnetic field lines and synthetic EUV emissions}
\label{app:euv_mag}

\begin{figure*}[htbp]
    \centering
    \includegraphics[width=\textwidth]{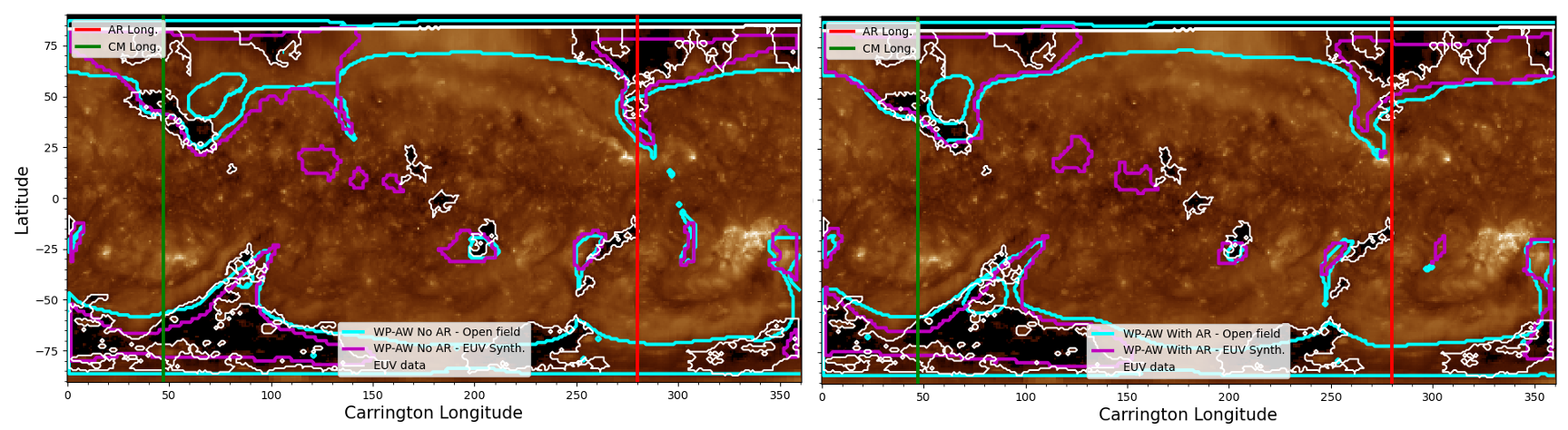}
    \caption{Comparison between coronal hole boundaries derived from open magnetic field lines (cyan) and synthetic EUV emissions (magenta). The left panel corresponds to the WP-AW simulation without the far-side AR (February 15), while the right panel corresponds to the WP-AW simulation with the far-side AR (February 22). Contours of the coronal holes extracted from the background SDO/AIA 193\,\AA\ EUV synoptic map are shown in white. We also indicate the longitude corresponding to the AR on the far-side with a red line, and the longitude corresponding to the central meridian as seen from the Earth with a green line. The comparison is made with CR2240.}
    \label{fig:ch_euv_mag}
\end{figure*}

As shown in section \ref{sec:results_ch}, there are different ways to estimate the boundaries of a coronal hole: one can consider that it is a region of open magnetic field lines, but one can also consider that it is a region of reduced EUV emission. These two definitions usually refer to the same regions, but it is not always the case, especially in numerical simulations.

Because of this, we have added in this appendix in Figure~\ref{fig:ch_euv_mag} the comparison between coronal hole boundaries derived from open magnetic field lines (cyan) and synthetic EUV emissions (magenta). The left panel corresponds to the WP-AW simulation without the far-side AR (February 15), while the right panel corresponds to the WP-AW simulation with the far-side AR (February 22). Contours of the coronal holes extracted from the background SDO/AIA 193\,\AA\ EUV synoptic map are shown in white. We also indicate the longitude corresponding to the AR on the far-side with a red line, and the longitude corresponding to the central meridian as seen from the Earth with a green line. The comparison is made with CR2240. The WP-AW simulations are the only ones for which we can perform such a comparison, as other models do not have a realistic enough corona to generate realistic EUV synthetic emissions.

From Figure~\ref{fig:ch_euv_mag}, one can see indeed that the two definitions overlap most of the time but not always. We observed that for both simulations (with and without the far-side AR), the equatorial CHs boundaries match the EUV data boundaries better with the synthetic EUV emission rather than the open magnetic field lines. Similarly, the polar CHs boundaries are less extended in longitude when using the synthetic EUV emission. This comparison may also hold for the real data, as the polar regions are more prone to line-of-sight effects that tend to artificially enhance the EUV emission. 

\section{Detailed description of the numerical models}
\label{app:codes}

\subsection{Potential-field source surface extrapolation}
\label{subsec:codes_pfss}

The PFSS model \citep{altschuler1969magnetic, Schatten1969, schrijver2003photospheric}, is an efficient means of extrapolating the coronal magnetic field from each GONG-ADAPT map in the absence of non-potential structures. Despite being an empirical model, it compares fairly well with MHD models \citep{Riley2006}. The PFSS model requires that the coronal magnetic field becomes radial at a given distance, called the source surface, and that there are no electric currents present in the coronal magnetic field (i.e., $\nabla\times B=0$). For this reason, PFSS models fail to capture highly twisted or stressed magnetic field structures in the low-corona (typically associated with eruptive activity), although they adequately recover the global structure of the corona and solar wind \citep[e.g.,][]{stansby2021active}. The ADAPT-GONG maps, provided in latitude-longitude ($\theta$-$\phi$) coordinates, are first broken down into spherical harmonics. We use a spherical harmonic basis formed from the Legendre polynomial functions $P_{lm}(\cos\theta)$, with degree $l$ and order $m$, given by
\begin{eqnarray}
Y_{lm}&=&c_{lm}P_{lm}(\cos\theta)e^{im\phi},\\ 
Z_{lm}&=&\frac{c_{lm}}{l+1} \frac{dP_{lm}(\cos\theta)}{d\theta} e^{im\phi}, \\ 
X_{lm}&=&\frac{c_{lm}}{l+1} P_{lm}(\cos\theta) \frac{im}{\sin\theta} e^{im\phi},
\end{eqnarray}
with the normalization of
\begin{eqnarray}
c_{lm}=\sqrt{\frac{2l+1}{4\pi}\frac{(l-m)!}{(l+m)!}}.
\end{eqnarray}
The photospheric field, $B_r(\theta,\phi)$, is decomposed such that the resulting coefficients $\epsilon_{lm}$ satisfy 
\begin{equation}
    B_r(\theta,\phi) = \sum_{l=1}^{l_{max}}\sum_{m=-l}^{l} \epsilon_{lm} Y_{lm}(\theta,\phi),
\end{equation}
where $\epsilon_{lm}$ is the strength of each spherical harmonic mode, extracted from the GONG-ADAPT maps \citep[see methodology in][]{finley2023accounting}. The coronal field is extrapolated using these coefficients following
\begin{eqnarray}
B_r(r,\theta,\phi) = \sum_{l=1}^{\infty}\sum_{m=-l}^{l}\alpha_{lm}(r)Y_{lm}(\theta,\phi),\\
B_{\theta}(r,\theta,\phi) = \sum_{l=1}^{\infty}\sum_{m=-l}^{l}\beta_{lm}(r)Z_{lm}(\theta,\phi),\\
B_{\phi}(r,\theta,\phi) = \sum_{l=1}^{\infty}\sum_{m=-l}^{l}\beta_{lm}(r)X_{lm}(\theta,\phi),
\end{eqnarray}
where $r$ is radial distance, and the coefficients $\alpha_{lm}(r)$ and $\beta_{lm}(r)$ represent the radial decay of each individual spherical harmonic mode: 
\begin{eqnarray}
\alpha_{lm}(r) = \epsilon_{lm}\frac{l(R_*/R_{ss})^{2l+1}(r/R_*)^{l-1}+(l+1)(r/R_*)^{-(l+2)}}{l(R_*/R_{ss})^{2l+1}+(l+1)},\\
\beta_{lm}(r) = (l+1)\epsilon_{lm}\frac{(R_*/R_{ss})^{2l+1}(r/R_*)^{l-1}+(r/R_*)^{-(l+2)}}{l(R_*/R_{ss})^{2l+1}+(l+1)}.
\end{eqnarray}
Each PFSS model is computed with $l_{max}=30$, on a spherical grid with source surface radii fixed at 2.5 solar radii. This value is the canonical value that matches the structures in coronograph observations for most phases of the solar cycle \citep{Hoeksema1983}. Some of the hypotheses can also be refined, such as using a non-spherical source surface \citep{Levine1982, Panasenco2020} or a complementary current sheet model \citep{Wang1995}. However, such refinements are beyond the scope of this paper, where we simply want to assess the quality of a standard PFSS extrapolation.


\subsection{Wind Predict}
\label{subsec:codes_wp}

We now present our first 3D MHD model WP. The model solves the set of the conservative ideal MHD equations composed of the continuity equation for the density $\rho$, the momentum equation for the velocity field $\boldsymbol{v}$ with its momentum written $\boldsymbol{m}=\rho\boldsymbol{v}$, the energy equation with $E$ the total energy and the induction equation for the magnetic field $\boldsymbol{B}$:
\begin{equation}
\frac{\partial}{\partial t}\rho+\nabla\cdot\rho\boldsymbol{v}=0,
\end{equation} 
\begin{equation}
\frac{\partial}{\partial t}\boldsymbol{m}+\nabla\cdot(\boldsymbol{mv}-\boldsymbol{BB}+
Ip) = \rho\boldsymbol{a},
\end{equation}
\begin{equation}
\label{eq:energy_wp}
\frac{\partial}{\partial t}E + \nabla\cdot((E+p)\boldsymbol{v}-\boldsymbol{B}(\boldsymbol{v}\cdot\boldsymbol{B})) = \boldsymbol{m}\cdot\boldsymbol{a},
\end{equation}
\begin{equation}
\frac{\partial}{\partial t}\boldsymbol{B}+\nabla\cdot(\boldsymbol{vB}-\boldsymbol{Bv})=0,
\end{equation}
where $p$ is the total pressure (thermal and magnetic, such that $p=p_{\rm{mag}} + p_{\rm{th}}$), $I$ is the identity matrix and $\boldsymbol{a}$ is a source term (gravitational acceleration in our case, which means $\boldsymbol{a} = -\nabla\Phi$ and the gravity potential $\Phi$ is equal to $-GM_\odot/r$). We also assumed $\nabla\cdot \boldsymbol{B} = 0$. We used the ideal equation of state:
\begin{equation}
\rho e = p_{th}/(\gamma -1),
\end{equation}
where $p_{th}$ is the thermal pressure, $e$ is the internal energy per mass and $\gamma$ is the adiabatic exponent. This gives for the energy : $E = \rho e+\boldsymbol{m}^2/(2\rho)+\boldsymbol{B}^2/2$.

PLUTO solves normalized equations, using three variables to set all the others: length, density and speed. If we note with $*$ the parameters related to the star (the Sun in this case) and with $0$ the parameters related to the normalization, we have $R_*/R_0=1$, $\rho_*/\rho_0=1$ and $u_{kep}/U_0=\sqrt{GM_*/R_*}/U_0=1$, where $u_{kep}$ is the Keplerian speed at the stellar surface and $G$ the gravitational constant. By choosing the physical values of $R_0$, $\rho_0$ and $U_0$, one can deduce all the other values given by the code in physical units. In our set-up, we choose the following reference solar values: $R_0=R_\odot=6.96 \ 10^{10}$ cm, $\rho_0=\rho_\odot=1.67 \ 10^{-16} \ \mathrm{g/cm}^3$ (which corresponds to the density in the solar corona above 2.5\,Mm, cf. \citealt{Vernazza1981}) and $U_0=u_{kep,\odot}=4.37 \ 10^2$\,km/s. 

One of the most critical aspect of coronal modeling is handling the coronal heating. In WP, we use the polytropic approximation. This means that we use a polytropic index equal to the adiabatic exponent $\gamma$ defined such as: $p \sim \rho^\gamma$. With such an approximation, we end up with values of $\gamma$ close to 1, which mimics a quasi-isothermal heating. 
Choosing $\gamma \neq 5/3$ is a simplified way of taking into account heating and thermal conduction, which are not modeled here. It is not possible to reproduce the bimodal distribution of the solar wind with this approximation \citep{Hazra2021}. However, it can successfully fit solar wind speed when we isolate the various populations \citep{Dakeyo2022, Shi2022}.

Our wind simulations are then controlled by four parameters: the normalized density $\rho_s$, the adiabatic exponent $\gamma$ for the polytropic wind, the rotation of the star normalized by the escape velocity $u_{rot}/u_{esc}$ and the speed of sound normalized also by the escape velocity $c_s/u_{esc}$. We note that the escape velocity is defined as $u_{esc} = \sqrt{2}u_{kep} = \sqrt{2GM_*/R_*}$. For the rotation speed, we take the solar value, which gives $u_{rot}/u_{esc} = 2.93 \ 10^{-3}$. For the density, we choose a ratio of $\rho_s=1$. We choose to fix $c_s/u_{esc}=0.26$, which corresponds to a $1.6 \ 10^6$ K hot corona for solar parameters and $\gamma=1.03$. 

We use the spherical coordinates $(r,\theta,\phi)$. The numerical domain is a sphere with the colatitude $\theta \in [0,\pi]$, the longitude $\varphi \in [0,2\pi]$ and the radius $r \in [1,30]R_\odot$. We use a uniform grid in latitude with 86 points and in longitude with 172 points, and a stretched grid in radius with 256 points; the grid spacing is geometrically increasing from $\Delta r/R_*=0.001$ at the surface of the star to $\Delta r/R_*=0.74$ at the outer boundary. 

Since PLUTO is a multi-physics and multi-solver code, we choose a finite-volume method using an approximate Riemann Solver (here the HLLD solver which stands for Harten-Lax-van Leer Discontinuities, cf. \citealt{Einfeldt1988}). PLUTO uses a reconstruct-solve-average approach using a set of primitive variables $(\rho,\boldsymbol{u},p,\boldsymbol{B})$ to solve the Riemann problem corresponding to the previous set of equations. 
We set the spatial order of integration to second order accurate, using a piecewise TVD linear reconstruction applied to primitive variables.
We use the monotonized central difference limiter (which is the least diffusive offered in PLUTO). 
We use the splitting-field option in PLUTO, which causes a splitting between the magnetic background field (which is curl-free) and the deviation field (which is a perturbation of the background field and carries the magnetic energy). To enforce the divergence-free property of the field, we use a hyperbolic divergence cleaning, which means that the induction equation is coupled to a generalized Lagrange multiplier in order to compensate the deviations from a divergence-free field \citep{Dedner2002}. The default value of $0.1$ is used for the parameter $\alpha$ which controls the rate at which monopoles are damped.

At the latitudinal boundaries ($\theta=0$ and $\theta=\pi$), we set axisymmetric boundary conditions. At the top radial boundary ($r=30 R_*$), we set an outflow boundary condition which corresponds to $\partial/\partial r=0$ for all variables, except for the radial magnetic field where we enforce $\partial(r^2B_r)/\partial r=0$. Because the wind has opened the field lines, this ensures the divergence-free property of the field. 
At the bottom radial boundary ($r=R_*$), we set a condition to be as close as possible to a perfect rotating conductor. This means that we set the electric field to 0 inside the star by aligning the poloidal velocity on the poloidal magnetic field. For the toroidal component, we set its radial derivative constant \citep[similar to][]{Matt2012}: $\partial^2 B_{\varphi}/\partial r^2 = 0$. All other quantities are kept constant, fixed at their initial values inside the star. The Lagrangian multiplier from the divergence cleaning method is forced to be constant across the boundary.
We use the input magnetic map as the bottom boundary condition for the radial magnetic field. We pass it on to the code by decomposing it on spherical harmonics, and then reconstructing it internally. We use a maximum degree of $\ell_{max}=30$ for the reconstruction. We also need to avoid the instability regime of the code where regions with $v_A/v_{esc}>3$ will cause negative pressures ($v_A$ being the Alfvén speed). This means that we may need to apply a global scaling factor to the input map in order to get convergence. In this case, the factor is 0.3. We could also change the input density of the simulation, but tests have shown that it changes more significantly the resulting structures and makes comparisons with other codes more challenging.

We initialize the velocity field with a polytropic wind solution and the magnetic field with a potential extrapolation of the field based on the input map. We then let the simulation relax for 200 normalized times to reach a converged steady state. 

\subsection{Wind Predict with Alfvén waves}
\label{subsec:codes_wp-aw}

This extension of the WP model has been described extensively in \cite{Reville2020,Reville2021,Reville2022} and \citet{Parenti2022}. We focus here on the differences between the basic version of WP described above. This means that if we do not mention a specific feature, then it is the same as in the WP version.

The addition of Alfvén waves (AW) aims at modeling more accurately the coronal heating by providing the additional source of energy and momentum necessary to create the observed wind bimodality. This is achieved through both the turbulence dissipation and the wave pressure, which provides additional acceleration to the solar wind. The gamma index is now set to $5/3$, which is the ratio of specific heat for a fully ionized hydrogen gas. 

In this set-up, on top of the previous physical quantities, we propagate two populations of parallel and antiparallel Alfvén waves from the boundary conditions. We can thus define the Elsässer variables as
\begin{equation}
    \boldsymbol{z}^{\pm} = \delta\boldsymbol{v}\mp \rm{sign}(B_r)\frac{\delta\boldsymbol{b}}{\sqrt{\mu_0\rho}},
\end{equation}
so that the sign $+$ ($-$) corresponds to the forward wave in a $+$ ($-$) field polarity. We can also define the corresponding wave energy density:
\begin{equation}
    \varepsilon^{\pm} = \rho\frac{|z^{\pm}|^2}{4},
\end{equation}
as well as the corresponding dissipation terms:
\begin{equation}
    Q_w^{\pm} = \frac{\rho}{8}\frac{|z^{\pm|^2}}{\lambda}\left(\mathcal{R}|z^\pm| + |z^\mp|\right),
\end{equation}
where $\lambda$ is the dissipation length scale set to $\lambda=\lambda_\odot \sqrt{B_0/|B|}$ ($\lambda_\odot$ is the correlation length at the base of the domain set to $15,000\rm{km}$, which is close to the size of supergranules, and $B_0 = 1$\,G), and $\mathcal{R}=0.1$ is a reflection coefficient used to generate turbulent dissipation in open magnetic field regions. 
The wave energy propagation follows the Wentzel-Kramers-Brillouin theory (WKB), which gives the following equation, solved alongside the MHD ones:
\begin{equation}
    \frac{\partial \varepsilon^\pm}{\partial t} + \nabla\cdot\left([\boldsymbol{v}\pm\boldsymbol{v_A}]\varepsilon^\pm\right) = -\frac{\varepsilon^\pm}{2}\nabla\cdot\boldsymbol{v}-Q_w^\pm,
\end{equation}
where $\boldsymbol{v} \pm \boldsymbol{v}_A \equiv \boldsymbol{v}_g^\pm $ is the group velocity of Alfvén wave packets.

As a result of these additional terms, the energy equation \ref{eq:energy_wp} is modified as follows:
\begin{equation}
    \frac{\partial}{\partial t}(E+\varepsilon+\rho\Phi) + \nabla\cdot\left[(E+p+\rho\Phi)\boldsymbol{v} - \boldsymbol{B}(\boldsymbol{v}\cdot\boldsymbol{B}) + \boldsymbol{v}_g^+\varepsilon^+ + \boldsymbol{v}_g^-\varepsilon^-\right] = Q,
\end{equation}
where $Q$ is a source term that we describe in more detail in the next paragraph.

The source term $Q$ added to the energy equation can be decomposed into three components:
$Q = Q_h - Q_c - Q_r$,
where $Q_h$ is an additional source of heating, $Q_c$ is the thermal conduction and $Q_r$ models thin radiative cooling. The $Q_w$ term, corresponding to the sum of the two dissipation terms for both wave populations ($Q_w = Q_w^+ + Q_w^-$), does not appear explicitly here as a heating term because we solve for the total (wave + fluid) energy equation \citep[see the erratum of][]{Reville2020}. The wave heating remains however the main source of heating in the domain (about 90\%).

$Q_h$ is an ad hoc function modeling chromospheric or coronal processes that would be different from waves, which can be written as
\begin{equation}
    Q_h = \frac{F_h}{H}\left(\frac{R_\odot}{r}\right)^2\rm{exp}\left(-\frac{r-R_\odot}{H}\right),
\end{equation}
where $H\sim 1R_\odot$ is the heating scale height, and $F_h$ is the energy flux from the photosphere (equal to $2\times10^4 \ \rm{erg}.\rm{cm}^{-2}.\rm{s}^{-1}$) {\citep{Reville2020}}.

The term $Q_r$ corresponds to an optically thin radiation cooling prescription, such as
\begin{equation}
    Q_r = n^2\Lambda(T),
\end{equation}
where $n$ is the electron density, $T$ is the electron temperature, and $\Lambda(T)$ is a function that follows the prescription of \cite{Athay1986}.
The term $Q_c$ corresponds to the thermal conduction, written as
\begin{equation}
    Q_c = \nabla\cdot(\alpha\boldsymbol{q}_s + (1-\alpha)\boldsymbol{q}_p),
\end{equation}
where $\alpha=1/(1+(r-R_\odot)^4/(r_{coll}-R_\odot)^4)$ creates a smooth transition between the collision and collisionless regimes at a specific height ($r_{coll}=5R_\odot$), $\boldsymbol{q}_s=-\kappa_0T^{5/2}\nabla T$ is the usual Spitzer-Härm collisional thermal conduction term ($\kappa_0=9\times10^{-7}$ cgs), and $\boldsymbol{q}_p = 3/2p_{th}\boldsymbol{v}_e$ is the electron collisionless heat flux prescribed in \cite{Hollweg1986}. 

Another important modification is that, following \cite{Reville2021}, the model now includes the transition region. This means that most of the values for the inner boundary conditions have to be changed. As a result, the chromospheric temperature is set to $2\times10^4$ K, and the density to $2\times10^{10} \ \rm{cm}^{-3}$. This also affects the new control parameters of the simulation. The transverse velocity perturbation parameter, which controls the amplitude of the Alfvén waves launched in the domain, is set to $\delta v=12 \ \rm{km}.\rm{s}^{-1}$. 

The equations are solved using a HLL solver with parabolic reconstruction method and minmod slope limiter \citep{Einfeldt1988}.
The non-solenoidal condition on the magnetic field is ensured through the constrained transport method instead of the divergence cleaning method. The equations are solved in the rotating frame, and thus Coriolis and centrifugal forces are accounted for in the momentum and energy equations. The azimuthal speed is thus set to zero in the inner boundary condition.

The top boundary condition is now located at $30 \ R_\odot$ because it is the best compromise between the code accuracy and run time capabilities. For the same reason, the maximum degree of spherical harmonics reconstruction has been set to 25 instead of 30. The mesh is optimized in three parts: a uniform grid from 1 to 1.004 $R_\odot$ with ten points to model the transition region, then a first stretched segment with high resolution from 1.004 to 15 $R_\odot$ with 182 points to model accurately the solar corona, and then a coarser stretched segment from 15 to 30 $R_\odot$ with 64 points for the inner heliosphere. 
The grid has 160 points in latitude (and removes the 5 degrees closest to each poles to avoid numerical singularities) and 320 points in longitude, and is regular in both these directions.

\end{document}